\documentclass{ar2e}
\usepackage{graphicx}
\usepackage{natbib}

\usepackage{epsfig,rotating,amsmath,subfigure}
\usepackage[3D]{movie15}
\usepackage{hyperref}
\usepackage[UKenglish]{babel}

\def\gtapprox{$\stackrel{>}{_{\sim}}$}
\def\lessapprox{$\stackrel{<}{_{\sim}}$}

\def\arcmin {$^\prime$}

\def\Msun  {${\rm M}_\odot$}
\def\deg   {$^\circ$}

\def\kms   {\ km s$^{-1}$}

\begin{document}

\input epsf.tex    
\input epsf.def   

\input psfig.sty


\title{Gaseous Galaxy Halos}

\author{M.E. Putman,  J.E.G. Peek,  M.R. Joung
\affiliation{Columbia University}}

\begin{abstract}

 Galactic halo gas traces inflowing star formation fuel and feedback from a galaxy's disk and is therefore crucial to our understanding of galaxy evolution.  In this review, we summarize the multi-wavelength observational properties and origin models of Galactic and low redshift spiral galaxy halo gas.  Galactic halos contain multiphase gas flows that are dominated in mass by the ionized component and extend to large radii.   The densest, coldest halo gas observed in neutral hydrogen (HI) is generally closest to the disk ($<20$ kpc), and absorption line results indicate warm and warm-hot diffuse halo gas is present throughout a galaxy's halo.  The hot halo gas detected is not a significant fraction of a galaxy's baryons.   The disk-halo interface is where the multiphase flows are integrated into the star forming disk, and there is evidence for both feedback and fueling at this interface from the temperature and kinematic gradient of the gas and HI structures.  
 \\
 The origin and fate of halo gas is considered in the context of cosmological and idealized local simulations.   Accretion along cosmic filaments occurs in both a hot ($>10^{5.5}$~K) and cold mode in simulations, with the compressed material close to the disk the coldest and densest, in agreement with observations.  There is evidence in halo gas observations for radiative and mechanical feedback mechanisms, including escaping photons from the disk, supernova-driven winds, and a galactic fountain.  Satellite accretion also leaves behind abundant halo gas.  This satellite gas interacts with the existing halo medium, and much of this gas will become part of the diffuse halo before it can reach the disk.  The accretion rate from cold and warm halo gas is generally below a galaxy disk's star formation rate, but gas at the disk-halo interface and stellar feedback may be important additional fuel sources.  
 
\end{abstract}

\maketitle

\section{Introduction}

Halo gas connects the baryon-rich intergalactic medium (IGM), to the star-forming disks of galaxies.  It represents a galaxy's future star formation fuel and the result of galactic feedback processes.   A galaxy's halo gas is therefore a combination of new material from the IGM and satellite galaxies and recycled material from the disk.  The metallicities of the long-lived stars in the Galactic disk indicate the vast majority of the incoming fuel should be low metallicity, and therefore feedback material cannot be the dominant fuel at all redshifts \citep{chiappini01, larson72, fenner03, sommer03}.   In addition, star formation rates (SFRs) indicate that ongoing accretion is required for galaxies at a range of redshifts \citep{erb08,putman09, hopkins08}; the Milky Way is no exception with only $\sim5 \times 10^9$ \Msun~of fuel in the disk and a current SFR of 1-3 \Msun/yr \citep{chomiuk11,smith78,robitaille10}.   All new galactic fuel needs to pass through a galaxy's halo, where the gravitational potential will draw the gas to the disk and help maintain spiral structure by keeping the Toomre Q parameter low \citep{toomre90,toomre64}.

This review outlines the observational properties and potential origins of the multiphase halo medium in both the Milky Way and other spiral galaxy halos.
Though a multiphase halo was first introduced theoretically by \cite{spitzer56}, the observational relationship between the multiple phases has only become more apparent in the past decade.  
 Low redshift studies of halo gas are key, as this is where physical details can be resolved and the low density gas can be mapped in emission.   
We define halo gas to lie within the approximate virial radius ($\sim250$ kpc for the Milky Way) and beyond the star forming disk.    The latter boundary is often difficult to define and most likely somewhat artificial given the ongoing transition of material between the lower halo and the disk.  
This disk-halo interface is discussed at several points in the review since it represents a key transition point.  The total accretion rate at this interface is also an important quantity to determine, given questions about a galaxy's future star formation and evolution.  

There are several possible origins for halo gas and a mixture of origins is likely for a spiral galaxy like the Milky Way at low redshift.  In cosmological simulations of galaxy formation, a large amount of the halo gas originates from inflowing IGM along cosmic filaments.   These simulations also find galactic feedback provides enriched material to galaxy halos.   Finally, satellites contribute gas to galaxy halos as they accrete onto the galaxy and are stripped of their gas.    
As we are now able to determine the physical properties of much of the multi-phase halo medium (with distance determinations and detailed observations), origin models can be more tightly constrained.
In addition, simulations are increasingly able to include the relevant gas physics and guide future observations.

Through observations and simulations we have made substantial progress in understanding gaseous halos and the flow of baryons in the universe. 
We begin this review with an overview of observations of the gaseous Galactic halo (\S\ref{sec:ghalo}).  Due to its proximity, this is the halo for which we have the most data, and we discuss its gas in sub-sections that are largely defined by temperature.
The gas at the disk-halo interface of the Milky Way is discussed separately in \S\ref{sec:diskhalo}.  
The Milky Way has a total mass of $\sim2 \times 10^{12}$ \Msun \citep{reid09, kalberla07}, and ideally we would discuss only galaxies of this mass for comparison.  In reality, there is no perfect nearby Milky Way equivalent, or limited observations of the best examples, and so \S\ref{sec:other} summarizes the halo gas properties of a range of low redshift spiral galaxies.    We consider the origin of the observed halo gas in the context of simulations and theoretical models in \S\ref{sec:origins} and comment on the survival of halo clouds in \S\ref{sec:survive}.   Finally, the sources available and possible methods of feeding galaxy disks are presented in \S\ref{sec:feeding} and the outlook for the future is noted in \S\ref{sec:future}.

\section{The Galactic Halo}
\label{sec:ghalo}

This section focuses on the significant observational progress that has been made in studies of Milky Way halo gas in the past decade (see \cite{wakker97} for earlier work) and is divided into sub-sections based primarily on the temperature of the gas.  
The neutral hydrogen halo gas discussed in \S\ref{sec:hi} is at temperatures $<10,000$ K and detected through the 21-cm line in emission.  This is traditionally the primary tracer of Galactic halo gas, and we therefore separately discuss its physical properties (\S\ref{sec:phys}).  The warm and warm-hot gas is defined here as largely ionized with temperatures $>10^{4}$~K, but $<10^6$~K (\S\ref{sec:warm}).  This gas is detected in optical line emission (primarily H$\alpha$) and absorption lines in the optical and ultraviolet .  Finally, hot gas is at temperatures greater than $10^6$~K and detectable through x-ray observations close to the disk and indirect methods further out (\S\ref{sec:hot}).  The dust, molecules and metals in halo gas are discussed in \S\ref{sec:dmm}.

Distances are often an issue when defining gas as a halo medium.  There has been significant progress in this area (see \S\ref{sec:phys}), but in any case, it should be noted that halo gas has largely been defined based on its velocity not falling within the expected Galactic disk velocity in a given direction.   This definition may in some (relatively rare) cases pick up Galactic gas that has been recently imparted with a large amount of kinetic energy, and it definitely misses halo gas that overlaps with the velocity of the disk.   In general, the transition between halo and disk gas is blurry, and a continuum of gas between the disk and halo seems likely.  Halo gas is defined here to be at heights above the Galactic Plane (z-heights) of a few kpc and not part of the Galactic warp, and gas below this that shows some connection to the disk is discussed in the disk-halo interface section (\S\ref{sec:diskhalo}).

\begin{figure}
\includegraphics[trim=70 20 50 250, clip=true, angle=90, scale=1.2]{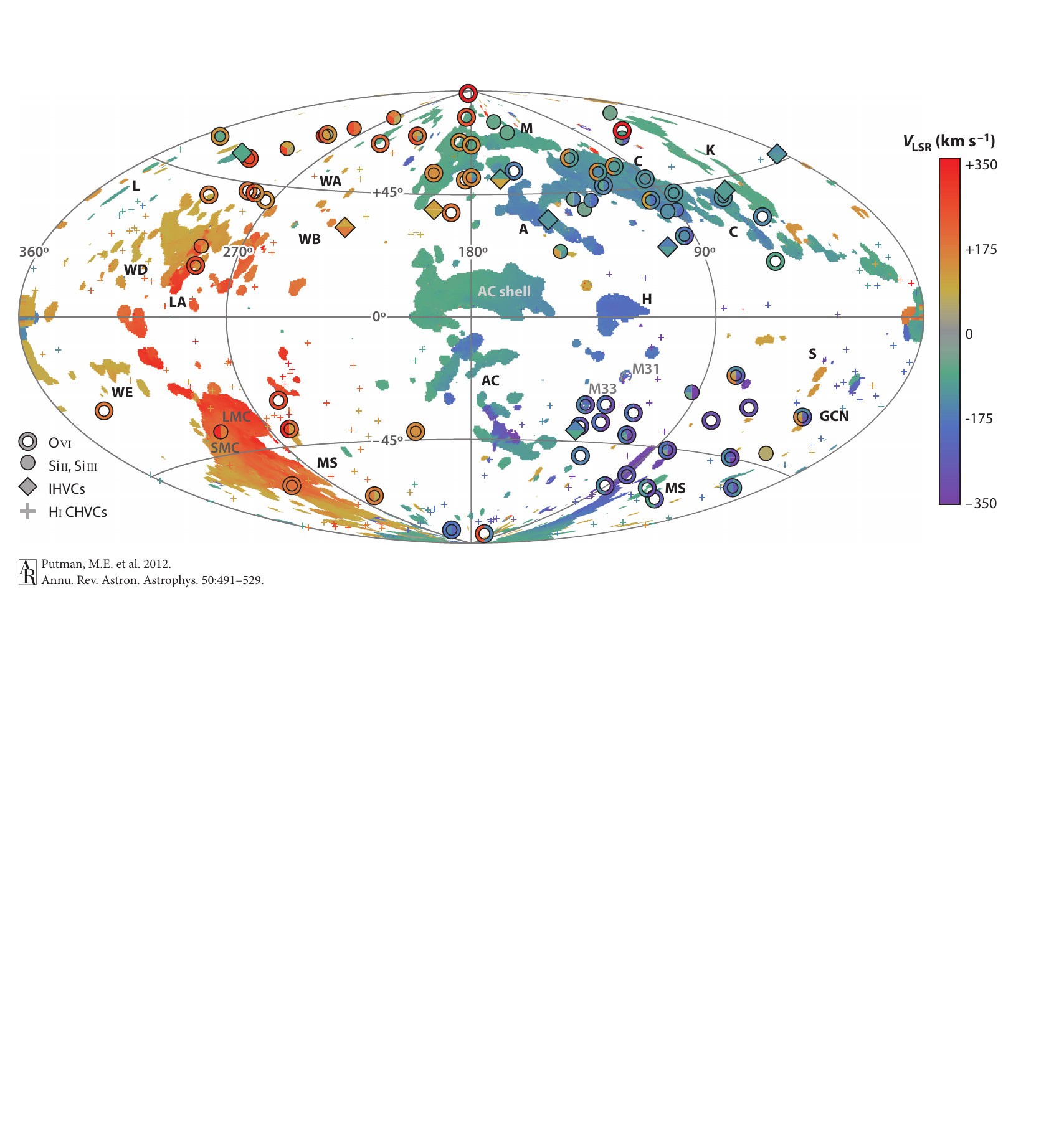}
\caption{\scriptsize The distribution of HI (shaded clouds and plus symbols) and ionized high velocity gas (circles and diamonds) on the sky with color denoting the LSR velocity of the detection. The HI map was created with LAB data by \cite{westmeier07t} by removing the HI model of the Milky Way from \cite{kalberla09}.
The plus symbols represent the the small HI compact HVCs detected with other datasets \citep{putman02,deheij02}, open circles are the O~VI absorption line detections \citep{sembach03}, solid circles are the Si absorption line detections \citep{shull09}, and diamonds are the ionized HVCs (IHVCs) at \lessapprox15 kpc \citep{lehner11}.  Multiple colors for a symbol indicate multiple absorbers along the line of sight.  The positions of the major HI HVC complexes (with S to indicate the Smith cloud) and the LMC, SMC, M31 and M33 galaxies are noted.  The AC Shell appears in this map, but is considered an intermediate velocity cloud.} \label{fig:assoc}
\end{figure}

\begin{figure}
\includegraphics[trim=75 170 50 170, clip=true, scale=1]{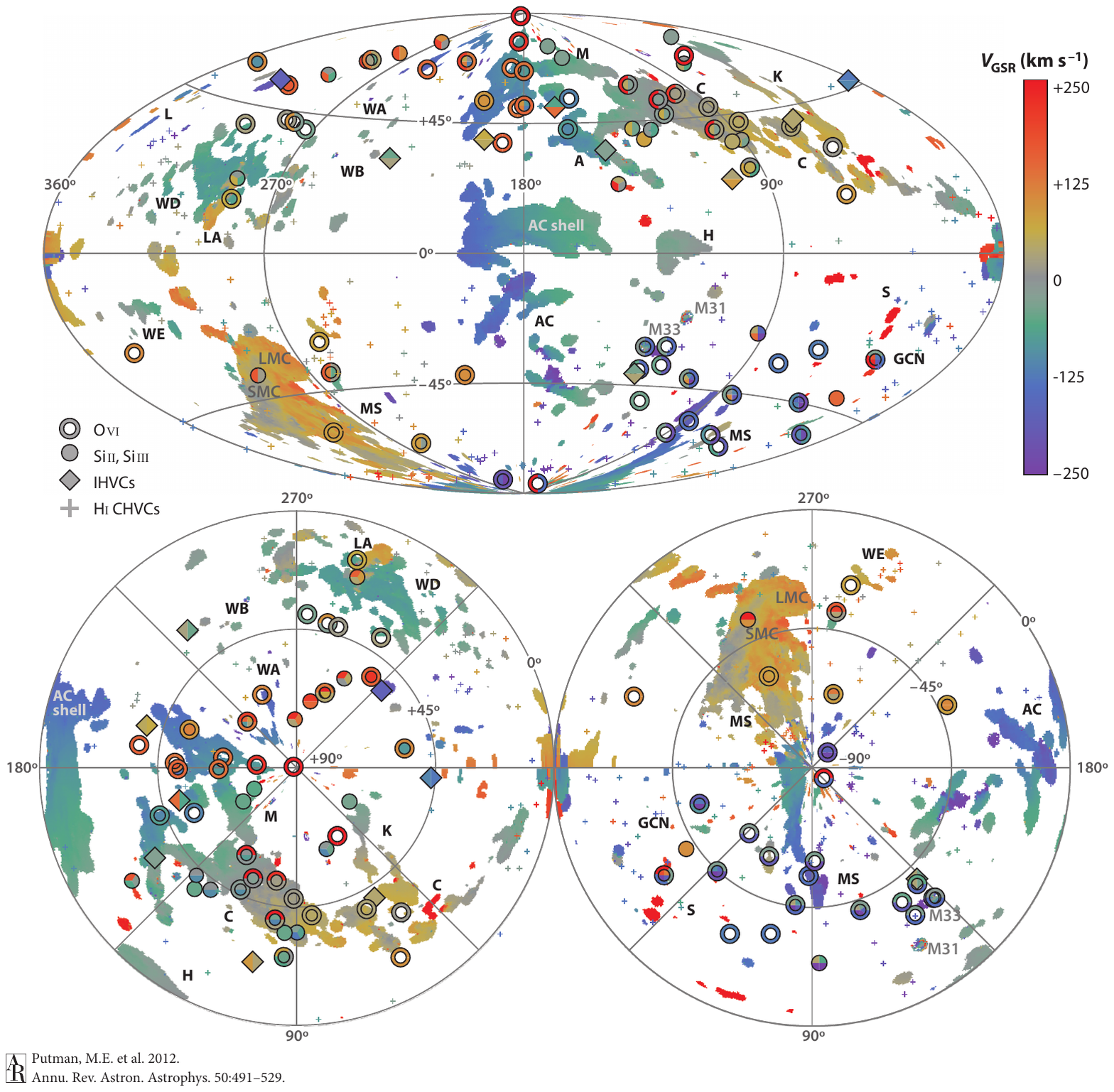}
\caption{\scriptsize The distribution of HI (shaded clouds and plus symbols) and ionized high velocity gas (circles and diamonds) on the sky with color denoting the GSR velocity of the detection.  Otherwise this figure is the same as Figure~\ref{fig:assoc}, with the exception of showing the view from the North (left) and South (right) Galactic Poles at the bottom. } \label{fig:assoc2}
\end{figure}

\subsection{Neutral Hydrogen Halo Gas}
\label{sec:hi}

The first halo gas detected was in neutral hydrogen emission at high velocity relative to Galactic disk velocities \citep{muller63}.  These discrete gas clouds were subsequently referred to as high-velocity clouds (HVCs).  HVCs are the coldest and densest component of halo gas and are found throughout the sky (Figures~\ref{fig:assoc} \& \ref{fig:assoc2}).   They are material stripped from satellites and/or cooling components of the accreting IGM and feedback material (see \S\ref{sec:origins}).  Significant progress in mapping the distribution of HVCs has come through multibeam instruments that have been used to observe large areas of sky at relatively high resolution \citep[4\arcmin~to 16\arcmin;][]{putman02, saul12, mcclure09}.  
In addition, the entire sky has now been mapped at a uniform resolution of 35\arcmin~and 1 km s$^{-1}$ 
with the Leiden/Argentine/Bonn (LAB) HI survey \citep[][; see Figures~\ref{fig:assoc} \& \ref{fig:assoc2}]{kalberla05}. The full-sky successor to the LAB survey will be a combination of multibeam surveys with a resolution of $9-16$\arcmin~and 1\kms~\citep{mcclure09,winkel10}.

HVCs range in size from 1000 deg$^2$ to the resolution limit of the observations, and large clouds are often found to be composed of a number of smaller clouds when observed with better resolution and equivalent sensitivity \citep{putman02, stanimirovic08, hsu11}. There are many
catalogs of clouds that have been created with the various HI survey data that use different catalog methods.  The number of individual clouds in these catalogs ranges from $\sim600$ over the entire sky in the lowest resolution surveys \citep[35\arcmin~beam;][]{wakker91}, to 2000 in just the southern sky \citep[15.5\arcmin~beam;][]{putman02}, to $\sim700$ clouds smaller than 20\arcmin~ in 7600 deg$^2$ of the sky \citep[4\arcmin~beam;][]{saul12}.  It remains difficult to separate the Galactic disk gas from halo gas. One method of cataloguing distinct halo features is to run algorithms on the entire HI data cube to determine structures that are distinct from each other in position-velocity space \citep[Maller et al. in prep.;][]{sharma09}.  

Clouds of all sizes can be grouped into complexes of most likely related clouds based on their spatial and kinematic proximity as labeled in Figures~\ref{fig:assoc} and \ref{fig:assoc2}.  In the Southern hemisphere the HVCs that dominate in sky coverage and mass are associated with the Magellanic System.   In particular, the complexes labeled MS and LA are the trailing Magellanic Stream and Leading Arm \citep[recent references include: ][]{putmanmagrev, putman03, putman98, bruens05, westmeier08, stanimirovic02, stanimirovic08, nidever10}.  These complexes were created from the interaction of the Large and Small Magellanic Clouds (LMC and SMC) with each other and with the Milky Way.  The Magellanic Stream is a long, continuous structure with a well-defined velocity and column density gradient, while the Leading Arm is a collection of clouds throughout the region leading the Magellanic Clouds.
 In the Northern hemisphere there are a number of complexes of similar spatial size, but Complex C is the largest.  The best way to view the HI component of the complexes at their approximate distances (see \ref{sec:phys}) is through the three-dimensional Figure~\ref{fig:3d}.   
  In both hemispheres there are a number of small clouds  called compact HVCs (CHVCs) that have been claimed to be at large distance due to their size ($<2$\deg~ in diameter) and relative isolation \citep{braun99, deheij02}.  CHVCs can largely be associated with known HVC complexes in position-velocity space as shown by the plus symbols in Figures~\ref{fig:assoc} and \ref{fig:assoc2}, and are therefore unlikely to be at greater distance.  Even the ultra-compact HVCs (\lessapprox~20\arcmin) have been shown to be related to HVC complexes \citep{saul12}.  
None of the CHVCs or HVC complexes have a clear association with stellar features in the halo \citep{hopp07,siegel05}, excluding the association of MS and LA to the Magellanic Clouds.   

The majority of the kinematic range of the HVCs can be seen in Figures~\ref{fig:assoc} and \ref{fig:assoc2}.  The range has been extended further with new surveys and is currently approximately $-500$ to $+450$ \kms~in the Local Standard of Rest velocity frame (V$_{\rm LSR}$) or between $-300$ and $+300$ \kms~in the Galactic Standard of Rest frame (V$_{\rm GSR}$).   HVCs can be defined by the cloud's deviation velocity (V$_{\rm dev}$), or the amount the velocity of the cloud deviates from a simple model of Galactic rotation.   Figures~\ref{fig:assoc} and \ref{fig:assoc2} were created with $|V_{\rm dev}| > 75$ \kms~ and thus there are some small differences from maps created with the classical $|V_{\rm LSR}| > 70-90$ \kms~definition \citep[e.g.][]{westmeier07t,wakker91}. 
The clouds at both the positive and negative velocity extremes are largely associated with the Leading Arm and the tip of the Magellanic Stream, respectively.  
Since HVCs have not been catalogued in the velocity range obscured by Galactic emission, it is difficult to assess the dispersion of the population.  Using the GSR frame, an estimate for the velocity dispersion of the HVCs is $\sim$100-120~\kms.

The HVCs have typical linewidths of 20-30~\kms, corresponding to velocity dispersions of $\sigma = 8.5-13$\kms \citep{putman02, deheij02, kalberla06}.   This linewidth is indicative of a warm neutral medium for HVCs
at $\sim$ 9000 K \citep{hsu11}.   Many HVCs also show a narrow component in their velocity profiles corresponding to cold gas at \lessapprox~500 K \citep{kalberla06}.  The two-phase structure of HVCs is expected for clouds in pressure equilibrium with a diffuse hot Milky Way halo medium \citep[][\S\ref{sec:hot};]{wolfire95}.  It is possible that HVCs that do not show this two-phase structure are currently in a more turbulent environment, or are embedded in a lower pressure part of the halo.

HVCs have typical peak column densities of $\sim 10^{19}$ cm$^{-2}$, and a small number of clouds have peak column densities above $10^{20}$ cm$^{-2}$ with the 4-15.5\arcmin~resolution HI surveys \citep{putman02, hsu11}.   Additional dense cores with HI column densities $> 10^{20}$ cm$^{-2}$ have been detected within some HVCs with synthesis observations \citep[e.g.,][]{wakker02, bekhti06}.
The clouds are found to extend down in column density to the limits of the HI observations \citep[$> 10^{17}$ cm$^{-2}$;][] {lockman02,braun04}.  This suggests HI surveys may be detecting only a fraction of the area of many complexes, although they are likely to be detecting the majority of the HI mass \citep[e.g.,][]{Peek09}.  There is also evidence that the  clouds detected in HI emission are linked to lower column density absorption line systems \citep{bekhti12,murphy95,nidever10}. At $\sim7 \times 10^{17}$ cm$^{-2}$ in HI emission the covering fraction is $\sim37$\% of the sky \citep{murphy95,lockman02}.

\begin{figure}
\includegraphics[trim=120 200 50 200, clip=true, scale=1.1]{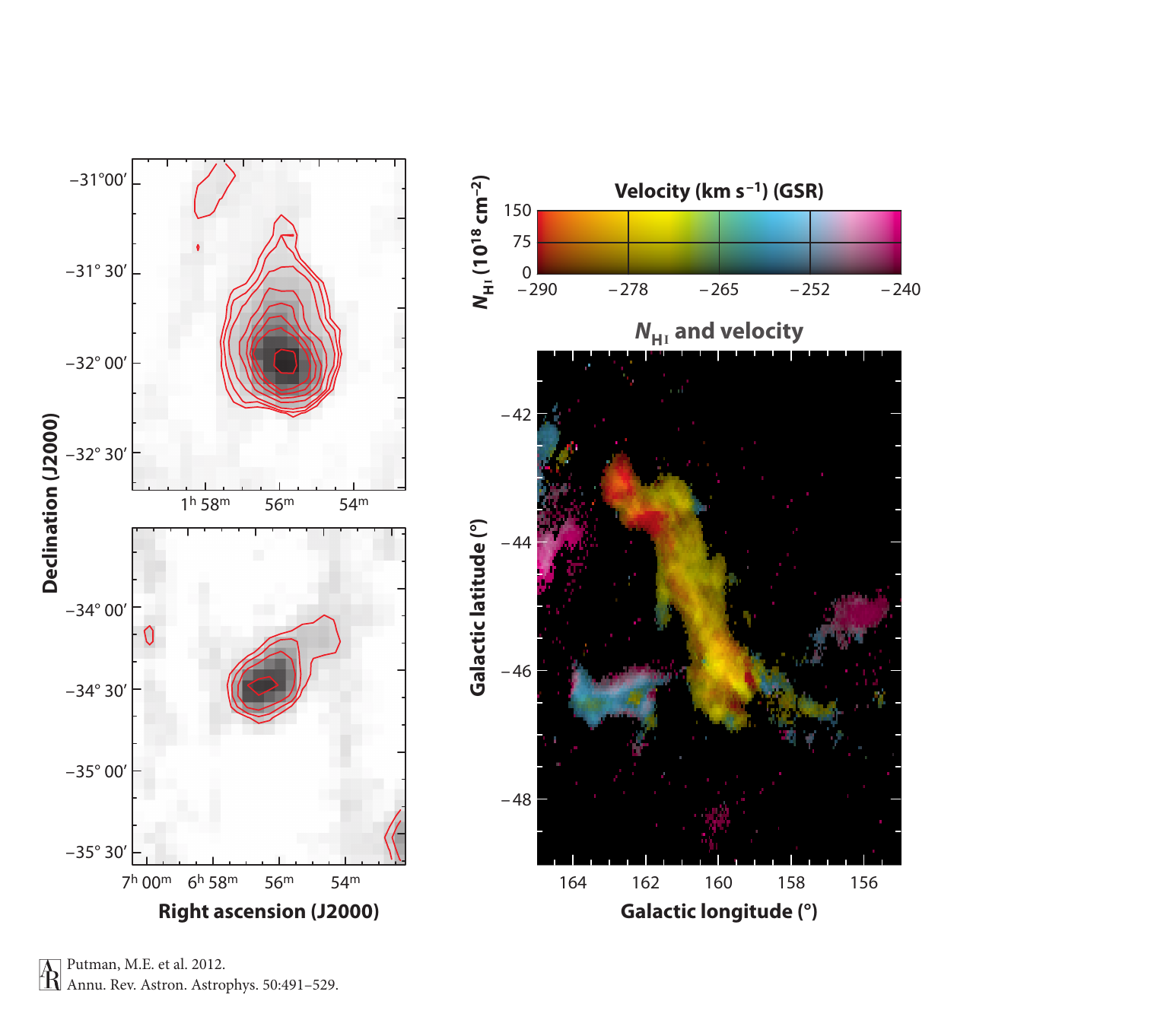}
\caption{\scriptsize Left:  Head-tail HI halo clouds with the outer contour at N$_{\rm HI} = 2 \times10^{18}$ cm$^{-2}$ \citep[15.5\arcmin~ resolution;][]{putman11}.    Right:  A HI HVC that shows signatures of interaction with the halo medium from the edge features \citep[4\arcmin~ resolution;][]{peek07}.}
\label{fig:ht}
\end{figure}

When the detailed HI structure of HVCs is examined, the clouds show evidence for being ionized by an external radiation field \citep{maloney03, sternberg02}.  The HI structure also often indicates the clouds are moving through a diffuse halo medium and being disrupted \citep[][see Figure~\ref{fig:ht} and \S\ref{sec:survive}]{peek07, bruens00, stanimirovic06}.  Head-tail clouds, that form as the leading side of the HVC is compressed and gas is stripped behind it, are particularly abundant in the vicinity of the Leading Arm \citep{putman11}.

\subsubsection{Distances and Physical Properties}
\label{sec:phys}

\begin{figure}
\includemovie[
	3Dviews2=views.tex,
	poster,
	toolbar,
	label=pt,
	text={\includegraphics[scale=0.45]{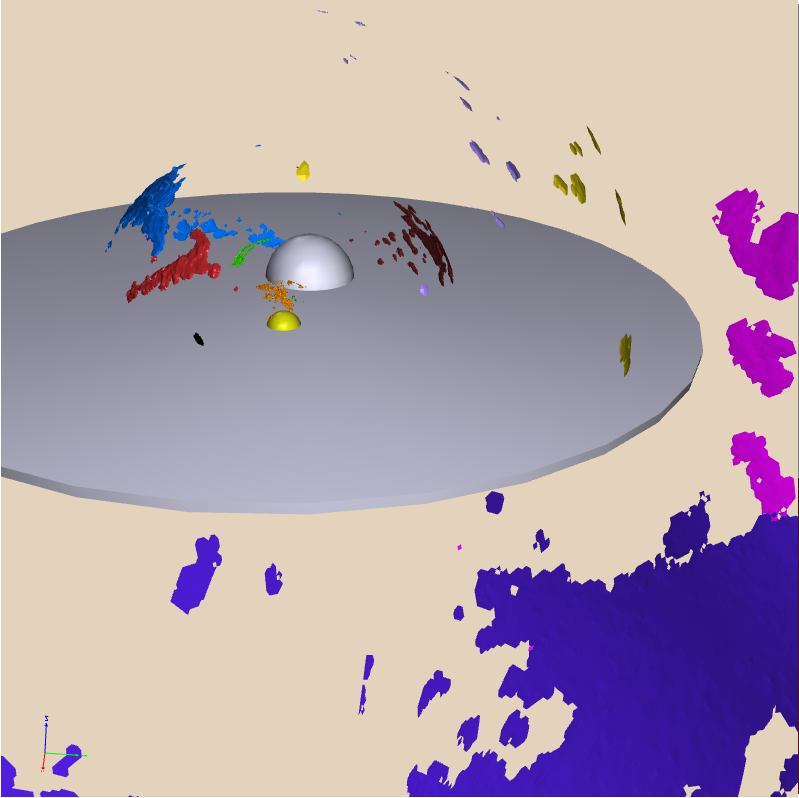}},	  
	3Dcoo=5.985595703125 6.663844108581543 -5.4499030113220215,  3Dc2c=0.8753247857093811 0.035229478031396866 0.482250452041626, 3Droo=43.90569672694108,  3Droll=-2.753936755660784,  3Daac=60.000001669652114,  3Dbg=0.894118 0.827451 0.737255,  3Dlights=CAD, 3Drender=Solid
	]{\linewidth}{\linewidth}{GalaxyModel.u3d}
	\caption{\scriptsize A 3D, interactive model of the Galactic HVC complexes. The Galactic disk is shown with a 25 kpc radius and a 1 kpc thickness. The central regions of our Galaxy are indicated with a gray 3 kpc radius sphere, and the Solar position is shown with a yellow sphere. HVC complexes are placed at the average of their most stringent distance constraints (\S\ref{sec:phys}).   The Magellanic Stream is shown starting at a distance of 50 kpc, and increasing linearly to 150 kpc at a Magellanic longitude of 150 degrees, consistent with the models of \cite{besla10}.  The positions of the LMC and SMC are noted with gray spheres. Cloud thickness is an arbitrary function of column density along the line of sight.  If the Figure does not appear interactive, please view with a current version of Adobe Reader and click and drag to rotate.}
	\label{fig:3d}
\end{figure}

HVC distance measurements allow for estimates of their physical properties such as mass, size, volume density, and pressure \citep[e.g.,][]{wakker01, hsu11}.    
The best method to measure the distance to a HVC is to use a halo star of known distance toward the cloud and search for absorption lines at the HVC's velocity in the star's spectrum.   A detection of absorption from the HVC shows that the cloud is in front of the star, while not detecting absorption may indicate the HVC is behind the star.   Usually the optical Ca II H and K and/or Na II doublet absorption lines from the HVC are used for this distance measurement.  Concluding the HVC is behind the star with a non-detection requires very deep observations.   This is because the line of sight to the star may have a lower HI column density than measured by the radio observations and there is uncertainty in the gas-phase abundance of the metals \citep{wakker01}.  The identification of halo stars with the Sloan Digital Sky Survey (SDSS) and other surveys has enabled distances to be obtained for almost all of the large, labeled HVC complexes in Figure~\ref{fig:assoc}.  Most of the complexes are at distances between 2-15 kpc and z-heights \lessapprox~10 kpc \citep[][Wakker et al., in prep.]{wakker08, thom06, thom08, wakker07}, and, as viewed from the Galactic Center, they are within 30\deg~of the disk plane.  The clear exceptions to these distances are the Magellanic Stream and Leading Arm, which are at $\sim55$ kpc near the Magellanic Clouds and may extend to 100-150 kpc at the tip of the Stream \citep[e.g.,][]{besla10}.  The complexes can be viewed at their distances in the three-dimensional map of Figure~\ref{fig:3d}.

Distance measurements for HVCs obtained with halo stars are referred to as direct distance constraints.  Indirect distance constraints are model dependent and available for almost all of the HVCs.   
H$\alpha$ observations of HVCs can be used to estimate their distance if one assumes that the H$\alpha$ emission is the result of the ionizing radiation from the Galaxy reaching the surface of the cloud \citep[the fraction escaping normal to the disk is f$_{\rm esc} \sim6$\% in this model;][]{blandhawthorn99, bland02, putman03b, bland01, bland98}.  Most of the distances derived from H$\alpha$ observations of HVCs have been consistent with their direct distance constraints using halo stars.  Clouds without direct distance constraints (in particular the CHVCs) are also usually detected in H$\alpha$ emission, which is consistent with them being within 40 kpc of the Galactic disk \citep{tufte02,putman03b}.  The striking exception to the escaping radiation model is the Magellanic Stream, which is too bright in H$\alpha$ emission given its distance (see \S\ref{sec:warm}).

The most important additional indirect constraint on HVC distances is deep HI observations of systems similar to the Milky Way and/or Local Group.  If one assumes the distribution of HVCs is similar, searches of Local Group analogs for HI clouds place even the compact HVCs within 80 kpc of the Galactic disk \citep{pisano07} and deep observations of Andromeda suggest the clouds are within 50 kpc \citep{westmeier07}.  The distance constraints have ruled out the possibility that the majority of the small HVCs are dark matter halos found throughout the Local Group \citep{blitz99,braun99}.  Recent Arecibo surveys are finding smaller clouds that do not yet have distance constraints \citep{giovanelli10, begum10}, but most of these clouds are associated with HVC complexes in position-velocity space \citep{saul12}.
 
The HI masses and sizes of HVC complexes can be determined once distances are known.  Calculating the HI mass requires a measurement of the total flux of the cloud and is proportional to the square of the distance.  
Determining the physical size of a HVC requires scaling the observed size directly with the distance, and this measurement may vary with the beam size and sensitivity of the observations.  The HI masses of the HVC complexes (each consisting of many individual clouds) are in the range of $10^5 - 5\times10^6$ \Msun, and their sizes are a few to 15 kpc across (excluding MS and LA).   There are now enough distance constraints that we can determine the total mass in halo clouds.  For the 13 complexes with direct distance constraints, we calculate the mass by placing each complex at the average of its upper and lower distance limits.   We place the remaining clouds ($\sim23$\% of the total HVC HI flux) at a distance of d=10 kpc, since the majority of the complexes are at approximately this distance and indirect distance constraints indicate there are not clouds at significantly larger distances.   This gives a total HI mass in non-Magellanic System HVCs of $2.6 \times 10^7$~\Msun.
This mass should be roughly doubled and multiplied by 1.4 to account for the ionized component (\S\ref{sec:warm}) and Helium respectively, thus giving a total mass of $7.4 \times 10^7$~\Msun.   The expected ongoing accretion rate from the HVC complexes is addressed in \S\ref{sec:accrete}.  The Magellanic System alone contributes a minimum of $3\times10^8$~\Msun (d/55~kpc)$^2$ just in HI to the halo from the Magellanic Stream, Leading Arm and Magellanic Bridge which joins the two Clouds \citep{putman03,bruens05}.  Including the LMC and SMC galaxies would add another $7 \times 10^8$ \Msun~in HI to the Milky Way halo.   All other satellites with HI gas are beyond the virial radius of our Galaxy \citep{grcevich09}.

The derivations of volume densities and pressures require additional assumptions about the HVCs and therefore should be adopted with some caution.  For volume densities, it is often assumed the cloud's path length is similar to the size of the dimensions observed.  For spherical clouds this may be a valid assumption if the linewidth does not seem unusually broadened.  Volume density estimates for clouds with direct distance constraints are $\sim0.05-0.15$ cm$^{-3}$, and when simulated halo clouds are 'observed' the measured volume densities are within a factor of 3-4 of their actual volume densities \citep{wakker01,hsu11}.  
An estimate of a HVC's pressure is made with the ideal gas law using the derived volume density and the kinetic temperature calculated from the linewidth (see \S\ref{sec:hi}).
Using the above range of densities and a temperature of 9000 K typical for the HVCs' warm neutral component, the pressures for the clouds with distance constraints are log(P/k) $= 2.7 - 3.1$ K cm$^{-3}$.   The HVCs are at z-heights from 3-9 kpc and the pressures derived are consistent with the calculations by \cite{wolfire95} for clouds in pressure equilibrium with a hot halo medium.   The estimates for the pressures associated with the Magellanic Stream clouds are substantially lower, and this would be expected for clouds in a more distant, lower density halo medium \citep{stanimirovic08, hsu11}.

\subsection{Warm Ionized and Warm-Hot Halo Gas}
\label{sec:warm}

The warm gas ($\sim10^{4-5}$ K) in the Galactic halo is detected with deep H$\alpha$ emission line observations and low and intermediate ion absorption lines  (e.g., Si II, Si III, C II, C III, O I) using background QSOs or stars.
 The H$\alpha$ observations are generally targeted toward known HI HVCs, while the location of the absorption line observations are dictated by the positions of background probes.  Both methods have led to numerous detections of warm Galactic halo gas.   This gas probes the interaction of HI clouds with photons from the disk and with other halo gas (\S\ref{sec:feedback}), and also comes directly from the IGM and/or satellites (\S\ref{sec:coldflow} \& \ref{sec:sat}).

There are numerous detections of HI halo clouds in H$\alpha$ emission with Fabry-Perot instruments and deep long-slit spectroscopy \citep{putman03b,tufte98, tufte02, bland98, weiner96, weiner02}.  All of the high-velocity H$\alpha$ detections are located on or near HI HVCs.  Extensions of the H$\alpha$ emission from the HI have been noted for one complex \citep{hill09}, but this does not appear to be a generic condition \citep{haffner05, haffner01, putman03b}.  The detections of H$\alpha$ emission are typically $< 0.5$ Rayleigh (R),
where 1 R is $10^6/4\pi$ photons cm$^{-2}$ s$^{-1}$ sr$^{-1}$.
There are some non-detections of H$\alpha$ emission for HVCs, but it is possible they will be detected in the future if the cloud is fully mapped to deep levels.  
One distinct non-detection may be the Leading Arm of the Magellanic System, which is not detected in H$\alpha$ for numerous pointings ($< 30-50$~mR).   This is in contrast to the Magellanic Stream, which is particularly bright in many locations and requires models that invoke collisional ionization mechanisms beyond photoionization by radiation from the disk \citep[][\S\ref{sec:phys} and \S\ref{sec:feedback}]{blandhawthorn07, weiner96, putman03b}.  
The velocities of the HI and H$\alpha$ lines are approximately the same for all HVC complexes, and there is no evidence for a correlation between HI column density and H$\alpha$ brightness.   Other species such as [NII] and [SII] have also been detected in emission for some HVCs \citep{bland98,tufte98, putman03b}.

The mass of the warm ionized component of halo clouds should be added to the neutral hydrogen mass to obtain the total mass in halo clouds.  The H$\alpha$ detections have been used to calculate that the ionized component of the HVC complexes is on the order of the mass of the neutral component in many cases \citep{hill09,wakker08, shull11}.  This suggests the warm ionized component of the HVCs contributes $3-4 \times 10^7$ \Msun~to the Galactic halo (excluding the MS and LA; see \S\ref{sec:phys}).  It should be noted that the ionized gas masses vary by significant factors depending on if one assumes the ionized and neutral gas are mixed together, or that the ionized gas forms a skin around the neutral component.  The numerous H$\alpha$ detections along the Magellanic Stream indicate this feature also has a significant amount of ionized gas associated with it \citep[see also][]{fox10}.

Absorption line observations are able to detect much lower column density gas at a range of temperatures compared to emission line observations.  An important conclusion from the absorption line observations is that multi-phase high-velocity gas covers a significant fraction of the sky.   A covering fraction of $\sim81$\% was found for T$\sim10^{4-4.5}$~K gas using high velocity Si~III lines \citep[$\langle$N$_{\rm Si~III} = 2.6 \times 10^{13}$ cm$^{-2}\rangle$;][]{shull09,collins09}, and a similar covering fraction ($>60$\%) was found for T$\sim10^{5-6}$ K gas using O~VI absorption \citep[N$_{\rm O~VI} > 2.5 \times 10^{13}$ cm$^{-2}$;][]{sembach03}.   These results can be used to infer that low column density ionized hydrogen covers the majority of the sky (see also Lehner et al. 2012\nocite{lehner12}).  Models show that the Si~III absorbers represent photoionized material, while the O~VI is largely from collisionally ionized gas produced as cool clouds interact with a hot halo medium \citep{sembach03}.  A combination of photo- and collisional ionization is commonly needed in models to reproduce the line ratios for all sightlines \citep{shull11,fox06, fox10, haffner11}.   \cite{lehner11} recently constrained the distance to a collection of ionized HVCs (IHVCs) using various ultraviolet absorption lines detected in the spectra of halo stars.  They found the ionized clouds with $|V_{LSR}| < 170$\kms~ are within $\sim15$ kpc and contribute on the order of $10^8$ \Msun~in gas to the halo (assuming a metallicity of 0.2 solar; see also Shull et al. 2009\nocite{shull09}).   Thus, the largely ionized component of halo gas most likely dominates in mass over the HI component, although one must be careful that the same velocity cuts are made with these comparisons.

In Figures~\ref{fig:assoc} and \ref{fig:assoc2}, high-velocity absorption lines are shown over the HI map of HVCs.  The absorption line sightlines were not chosen to lie near HI complexes \citep{sembach03, shull09}, but in these maps it can be seen that the majority of the absorbers are in the spatial and kinematic vicinity of the HI complexes.  
In particular, an extension of the Magellanic Stream is evident with absorbers at high negative velocities \citep[see also][]{nidever10,lockman02,gibson00}.  There are also extensions of Complexes C, A, and M and the W complexes, and many of the distance constraints for the ionized HVCs are consistent with those of the closest HI HVCs in position-velocity space.  This map clearly shows that large scale multi-phase flows are active in the halo.   There are some cases where the absorbers do not show a position-velocity link to complexes detected in HI.    Most of these are at positive velocities and $b>45$\deg.
These largely ionized structures could be longer extensions of the W and LA complexes, gas along local galaxy structures (i.e., toward the Local Supercluster in the north or toward the Sculptor Group for the positive velocity absorber near the South Galactic Pole), or outflows from the disk \citep[e.g.,][]{keeney06, zech08}.

\subsection{Hot Halo Gas}
\label{sec:hot}

The hot gas ($\sim10^6$ K) in a spiral galaxy halo traces feedback mechanisms (\S\ref{sec:feedback}), and potentially shock-heated IGM (\S\ref{sec:coldflow}).  The extent and mass of this medium is of great interest to cosmological models of galaxy formation, as they predict it extends out to the virial radius and potentially hosts a large percentage of the galaxy's baryons \citep{white78, maller04, fukugita06, kaufmann06, sommer06, crain10}.

A hot Galactic halo and/or disk-halo medium is inferred to exist from x-ray observations in emission \citep{kerp99,wang93} and absorption \citep[O~VII and O~VIII;][]{williams05,wang05}, and from dispersion measures of pulsars at a range of distances \citep{taylor93, gaensler08}.   Most of the hot gas detected directly in emission is thought to be within a few kpc of the disk \citep{fang06, yao07}, and this is consistent with other galaxies (see \S\ref{sec:otherhot}).  There is evidence for a lower density medium that extends out to at least the Magellanic System ($\sim50-100$ kpc) from several indirect methods.  1)  The abundant detections of high velocity O~VI absorption in the Galactic halo is attributed to the interaction of cool halo clouds with the hot halo medium \citep{sembach03}.  2)  Satellite galaxies are stripped of their gas out to large radii, most likely as they move through this diffuse hot halo medium \citep{grcevich09, blitz00}.  3)  The spatial and kinematic structure of the cold halo clouds indicate they are both supported and destroyed by a hot halo medium (see \S\ref{sec:survive}).  4) There is a low-density bipolar structure of unknown distance detected in x-ray \citep{bland03} and gamma-ray maps \citep{su10} that is consistent with hot gas being sent into the extended halo.

Mass estimates for the hot halo medium are $<10^{10}$ \Msun~ within 100 kpc of the disk \citep[e.g.,][]{collins05, anderson10, yao08}, and the covering fraction is estimated to be $>60$\% \citep{fang06, sembach03}.
As discussed further in \S\ref{sec:origins}, the amount of hot gas in the halo is below that predicted by cosmological simulations \citep[see also][]{bregman07}, but its extent makes it relevant to studies of large scale Galactic winds and/or shock-heated IGM.     The precise physical properties of the extended hot halo medium remain uncertain, though densities are generally determined to be between $10^{-5}$ and $10^{-4}$ cm$^{-2}$ at 50-100 kpc with the indirect methods mentioned above \citep{putman11,hsu11,stanimirovic06,grcevich09,sembach03}.  The temperature is usually assumed to be $1-2 \times 10^6$ K for these halo density estimates, consistent with the x-ray observations and with expectations for gas in equilibrium with a Milky Way mass halo.  This temperature also happens to be consistent with the velocity dispersion of the HI HVCs ($\sim$100-120 \kms).  There is a great deal of additional work required to understand this difficult-to-detect, low density, hot halo medium.  In particular, the distribution of this medium and its density at the virial radius need to be measured in order to obtain a more accurate estimate of its total mass.

\subsection{Dust, Metals and Molecules}
\label{sec:dmm}

In this section we discuss the observations of dust and metals in halo gas and the limited detections of molecules.  Halo dust and metals are direct tracers of the ejection of material from galaxies as they result from stellar evolution and supernovae.   Small dust grains are the dominant heating mechanism of the cold neutral medium and metals the dominant coolant, so their abundances dictate the temperature structure of neutral halo clouds \citep{wolfire95}.  Dust is destroyed by grain-grain collisions and by particle-grain collisions (sputtering) via shocks \citep{JT96, DS79b,DS79a}, and therefore halo dust properties can provide information on the dynamical history of the gas such as the launching mechanism.   
Understanding halo dust is also important for correcting the observed colors of extragalactic objects.  The standard Galactic extinction map was made using a single grain size and composition for all Galactic gas, and a single temperature along a given line of sight \citep{SFD98}.   Metallicity measurements can further help to determine the origin of the gas (see \S\ref{sec:origins}), while molecular gas may point to future star formation in the halo.

Dust can be detected in the Galactic halo through far-infrared (IR) emission, depletion of refractory elements, and the reddening of light from distant objects.  While far-IR emission provides a clean indication of dust in the disk, it is difficult to detect in the halo given the lower HI column densities and metallicities of halo clouds compared to the disk.   HVCs may have less of their metals in dust grains \citep{wakker01} and have cooler (and dimmer) dust.  No definitive detections of HVC dust have been made with far-IR emission, although there are tentative detections of Complexes M and C \citep{peek09b, planck11,MAMD05}. 
Measuring the depletion of refractory elements onto grains can in principle indirectly show the presence of dust, but it is difficult given the unknown ionization states of the gas across species, the likely variation in inherent abundance patterns for clouds, and even the significant disagreement in the baseline non-depleted solar abundance ratios \citep{asplund09}.  Nevertheless, there is no sign of the depletion of S and Si onto grains for Complex C \citep{collins07, tripp03} or ionized HVCs \citep{richter09}, while evidence for dust depletion has been found for clouds associated with the Magellanic System \citep{lu98, fox10}. 
Reddening of background sources is independent of temperature and ionization state and may be used in the future to detect or rule out dust in Galactic halo gas at interesting levels.  

Metallicity measurements trace the star-formation history of halo gas more directly than the dust.   Metallicities have been measured primarily only for the large Complex C and Magellanic System complexes (MS and LA; see Figures~\ref{fig:assoc} - \ref{fig:3d}), although additional estimates will be obtained with future Cosmic Origins Spectrograph (COS) observations \citep{wakker09hst}.  Metallicities for Complex C are 10-30\% solar, and there is some evidence for mixing with Galactic material at low latitudes \citep{shull11,collins07}.   The nearby Complex A has a similar measurement \citep{wakker01}.    The deuterium abundance was also measured for one sightline in Complex C and is consistent with cosmological expectations \citep[N(D~I) $= 2 \times 10^{15}$ cm$^{-2}$; $D/H=2.2 \times 10^{-5}$;][]{sembach04}. 
The Magellanic Stream and Leading Arm metallicity determinations are also $\sim10-30$\% solar \citep{fox10,gibson00,lu98,sembach01}.  This metallicity, and the relative abundances, have been used to support an SMC origin for MS and LA; however, the metal content of these features depends on when, and from what region, the gas was stripped from the Magellanic Clouds.

Molecular gas has only been detected along a few sightlines through HVCs.   H$_2$ was detected in the Leading Arm and Magellanic Stream \citep[$\sim10^{16-17}$ cm$^{-2}$;][]{sembach01, richter01}.   It was not detected toward a sightline through Complex C, but this may be due to a low HI column along the sightline \citep{richter03}.   CO and H$_2$ was detected in the bridge of gas connecting the LMC and SMC, which is consistent with the star formation in the Magellanic Bridge \citep{muller03, lehner02}.

\subsection{The Disk-Halo Interface}
\label{sec:diskhalo}

Observations of gas at the disk-halo interface help to determine the relationship between halo gas that comes from the disk and halo gas that fuels the disk.  
For the Milky Way, the gas at the interface between the disk and halo has traditionally been studied separately from the halo gas.  The notable exception is the hot gas because the velocity resolution of these data do not allow the two to be  separated.  Defining individual objects at the disk-halo interface is difficult at any wavelength.  Objects are defined with HI data with the large intermediate velocity clouds (IVCs) and a population of small discrete clouds, but at other wavelengths the properties of the entire gas layer are usually presented.

\begin{figure}
\includegraphics[trim=160 200 50 220, clip=true, scale=1.2]{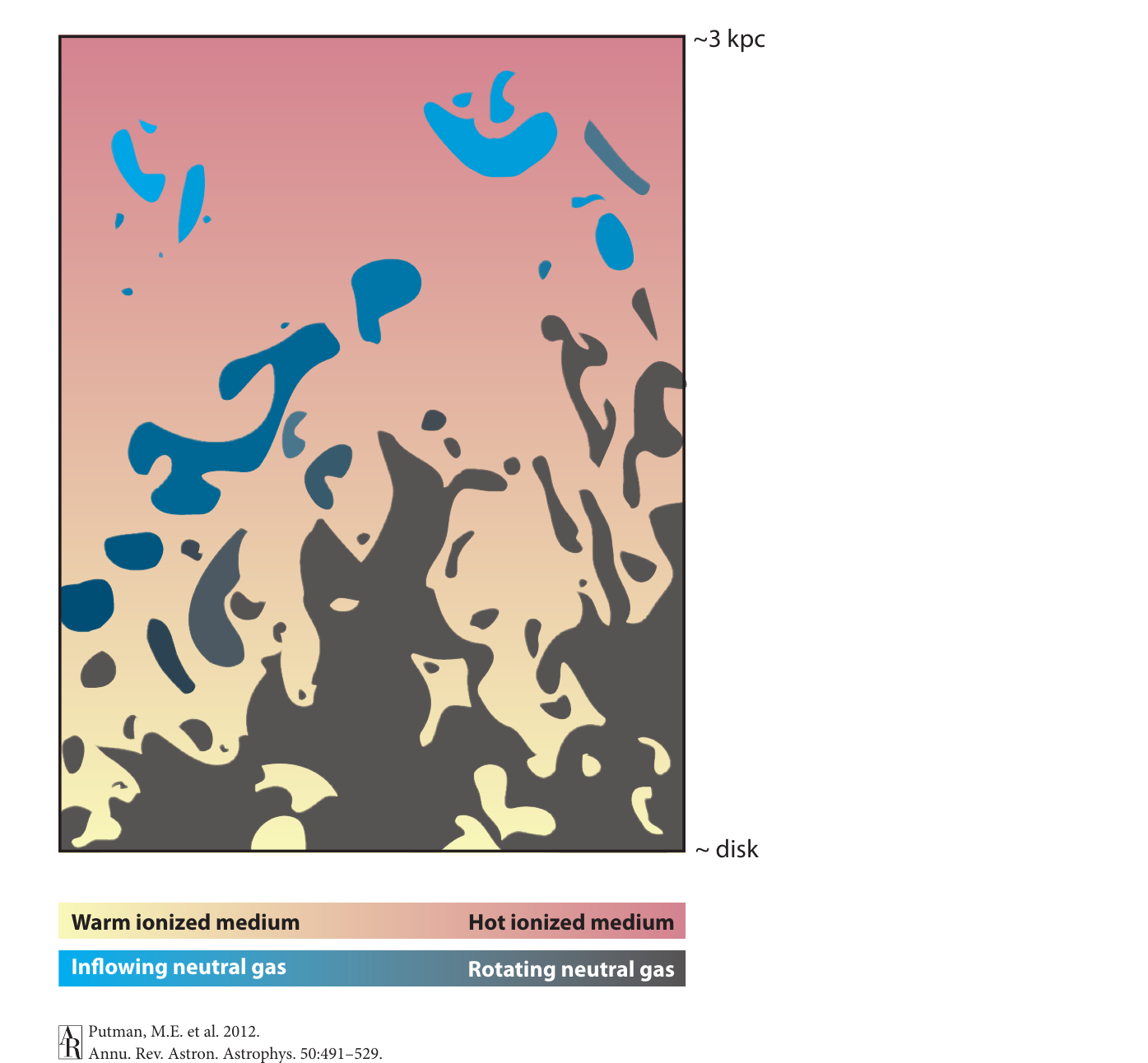}
\caption{\scriptsize A cartoon representing the disk-halo interface given the observations discussed in \S\ref{sec:diskhalo} and \S\ref{sec:other}.   The temperature gradient for the diffuse gas is shown with the pink color bar, while the kinematic gradient of the denser neutral gas (HI) is shown by the blue-gray shading.  There are distinct populations of large IVCs and small discrete clouds, but in reality the edges of the HI features are not this sharp.}\label{fig:diskhalo}
\end{figure}

For the cold gas, the thickness of the HI layer in the inner Galaxy is 220 pc, but approaches $<100$~pc in the nuclear regions and flares dramatically at R $>$ R$_\odot$ \citep[][see Figure~\ref{fig:other}]{dickey90}.  In some regions, there are kpc extensions from the disk due to the presence of IVCs or the Galactic warp \citep{levine06,kalberla08}.  
There is observational evidence that the HI temperature increases with increasing z-height.  This is also seen in the warm and warm-hot gas (see below).
Determining the variation in the movement of the gas with z-height is difficult given our location. The model of \cite{levine08} has a vertical falloff of the rotation curve of -22 $\pm$ 6 \kms~kpc$^{-1}$ near the disk, while \cite{marasco11} find lower values of -15 $\pm$ 4 \kms~kpc$^{-1}$.  Both the temperature and kinematic gradient of the disk-halo gas is depicted in Figure~\ref{fig:diskhalo}.

The large HI IVCs are not easy to isolate as individual objects due to their connection to Galactic HI.  They are found to reach z-heights of 1-2 kpc and their HI mass estimates are $0.5-8 \times 10^5$ \Msun. IVCs have near-solar metallicities \citep{wakker01,wakker08}, and dust is readily detected through IR emission. IVCs tend to have somewhat smaller, hotter dust grains than the disk, consistent with dust shattering through shocks \citep{heiles88, peek09b, planck11}.  Small amounts of molecular gas have been detected in numerous IVCs with H$_2$ column densities of $10^{14-17}$ cm$^{-2}$, consistent with the presence of dust \citep{richter03}.  Deuterium has also been detected \citep{savage07}.

Recently a large number of small, discrete HI clouds have been identified as a distinct disk-halo population and modeled as clouds rotating with the Galaxy at z-heights of 1-2 kpc \citep{begum10,saul12,ford10,lockman02b,kalberla09}.   The GALFA-HI survey has classified the largest population of these clouds ($\sim1250$), and can split the discrete HI clouds at intermediate velocities (20-90 \kms) into a cold (T $< 5000$ K) population that appears to be relatively uniformly distributed and related to the disk, and a warm population that has a bias toward negative velocities and is distributed similarly to the large IVCs \citep{saul12}.    The HI mass of the clouds depends on their distance and range from $\sim1$ \Msun~ for clouds at high Galactic latitudes at a distance of 1-2 kpc \citep{saul12}, to $\sim700$ \Msun~ for clouds toward the inner galaxy with distances obtained with the tangent point method \citep{ford10}.  The metallicity and dust content of these clouds has not yet been measured.

The warm gas detected in H$\alpha$ emission at the disk-halo interface is referred to as the Reynolds or warm ionized medium (WIM) layer  \citep{haffner03, reynolds93}.   This is a layer of warm gas extending $\sim2$ kpc above the disk with a volume averaged density of $0.01 - 0.1$ cm$^{-3}$ and a filling factor of $>30$\% at 1-1.5 kpc \citep{reynolds91,gaensler08, haffner09}.   As illustrated in Figure~\ref{fig:diskhalo}, at lower z-heights more of the volume is filled in with HI and at higher z-heights the hot gas fills more of the volume \citep{savage09,gaensler08}.   The scale height of gas at multiple temperatures, as well as its relation to the Galactic disk, has been investigated via absorption line studies \citep{savage09, bowen08}.  The cool and warm components have scale heights calculated from the absorption line results that are consistent with that obtained from the emission line observations.  The warm-hot gas, that is often referred to as transition temperature gas, is observed with Si~IV, C~IV and O~VI absorption lines and has a larger scale height ($\sim$3-4.5 kpc) than the cold and warm gas.   This warm-hot halo gas is found toward the vast majority of the sightlines in the Galactic disk.  The disk-halo interface region is discussed further in \S\ref{sec:diskhalofeed}.

\section{Halos of Other Spiral Galaxies}
\label{sec:other}

Halo gas observations for other spiral galaxies provide insight into the uniqueness and location of the Milky Way halo gas and the likely origin of the gas.   Extragalactic observations are generally not as deep as Milky Way observations in the case of emission line observations, and limited to a single sightline for absorption line observations.   
For emission line observations, the break between what is disk gas and what is halo gas depends on the spatial and kinematic model of the galaxy adopted.   Because the halo gas usually connects to the disk gas in extragalactic observations, the break is often somewhat arbitrary and varies in the literature.   For absorption line observations, we are limited to the projected distance and velocity of the absorber as a means to associate the gas with the halo of a galaxy.  

 This section provides a brief overview of observations of halo gas in other galaxies with the same sub-divisions as \S\ref{sec:ghalo}.  The exception in this section is that there is no attempt to discuss disk-halo gas separately, since the physical resolution of extragalactic observations is generally poorer.
The spiral galaxies noted in this section are just a few examples, and though they are generally Milky Way-like, they do span a range of total mass and SFR.   Differences in intrinsic galaxy properties can lead to different expectations for the amount of halo gas (see \S\ref{sec:origins}).

\subsection{Neutral Hydrogen Halo Gas}
\label{sec:otherhi}

HI halo clouds have now been detected beyond the disks of numerous spiral galaxies.   The lowest HI mass clouds observed around other spirals is $\sim10^5$ \Msun~\citep{thilker04}, but smaller clouds are likely to be detected with deeper and higher resolution observations.   The detected clouds are generally within $\sim10$ kpc of the disk, and deep observations do not show evidence for HI clouds at distances $> 50-80$ kpc from the disks \citep{westmeier07, pisano07}.  \cite{sancisi08} find that extraplanar features with masses \gtapprox~$10^8$ \Msun~exist in the halos of $\sim 25$\% of the field galaxies.  $10^8$ \Msun~is a very large HI feature (on the order of the Magellanic Stream), and if the criteria for a HI feature to be distinct is relaxed to include warps and/or kinematic evidence for deviations from disk rotation, the detected fraction increases to $\sim$50\% \citep{sancisi08, haynes98}.  Thus far, the majority of the spiral galaxies observed to deep enough levels show a disk-halo component that lags the rotation of the disk \citep[often referred to as an `HI beard';][]{sancisi01} with a typical gradient of magnitude 15-30 \kms~kpc$^{-1}$ \citep{fraternali02, oosterloo07, heald11}.  These deep observations also often find very limited amounts of distinct halo gas \citep{irwin09,zschaechner11}.

\begin{figure}[h]
\includegraphics[trim=115 185 50 210, clip=true, scale=1.1]{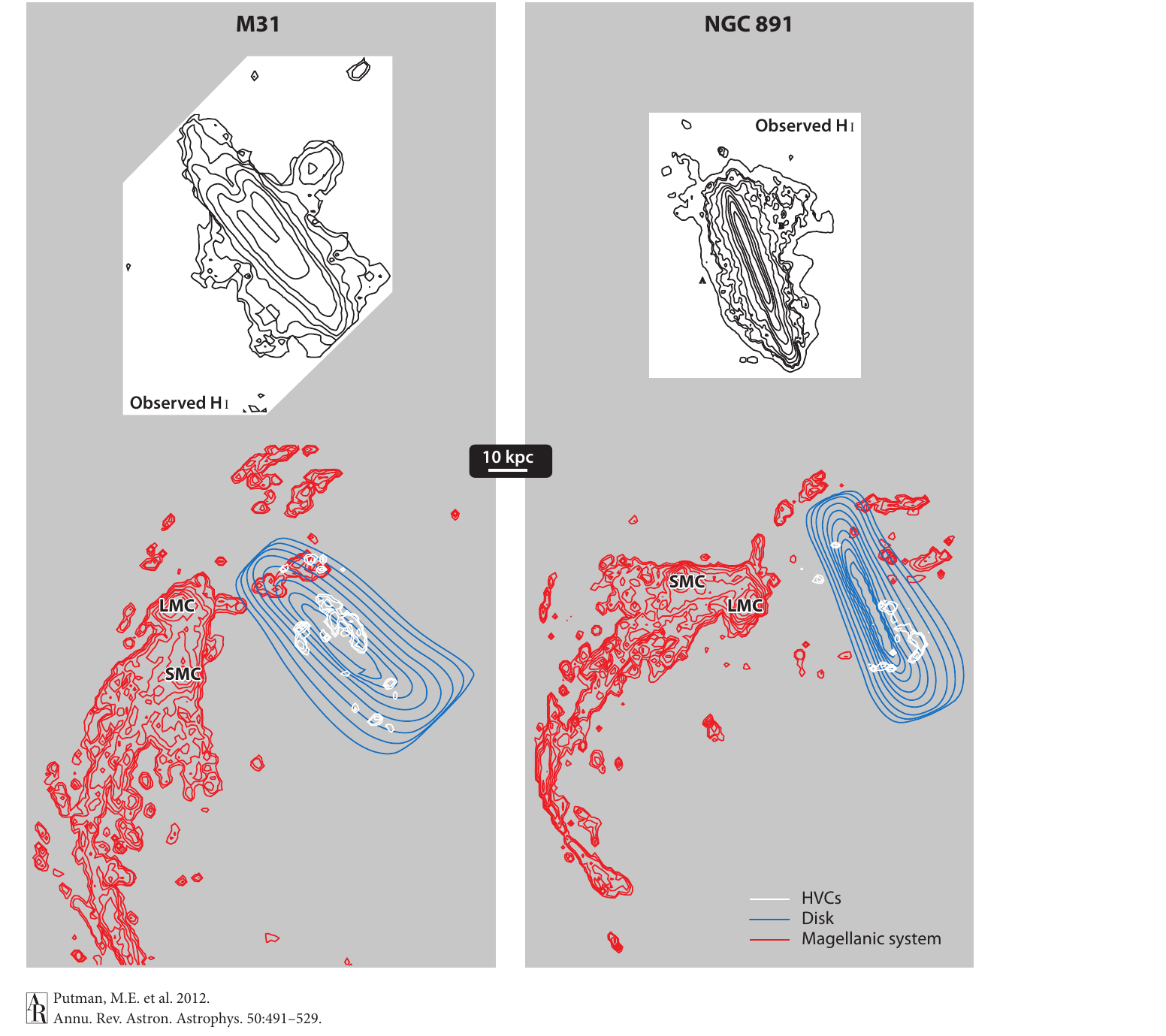}
\caption{\scriptsize Comparisons of HI halo gas for M31 (left) and NGC 891 (right) with that in the Milky Way if viewed with the same inclination. The exterior MW HI view is reconstructed using the distances to the 13 known HVC complexes (white) and assuming a distance for the MS/LA complex (red) ranging from 50 to 150 kpc using the models of \citep{besla10}. The MW disk (blue), including the disk column density, flare, and warp, is reconstructed from the work of \cite{levine06}. Contours are set to match the observations: 0.1, 0.3, 1, 3.2, 10, 32, 100, 316, 1000 x 10$^{19}$ cm$^{-2}$ for the \cite{westmeier07} M31 data and 0.5, 1, 2, 5, 10, 20, 50, 100, 200, 500 x 10$^{19}$ cm$^{-2}$ for the \cite{oosterloo07} NGC 891 data, respectively.  The large cloud above M31's disk is unlikely to be associated with M31 \citep{westmeier07}, and see \cite{ibata07} and \cite{mouhcine10} for stellar features around M31 and NGC 891, respectively.  Viewed externally, the Magellanic system is clearly the dominant feature in our HI halo and our HI disk is comparable to M31 and NGC 891.  We find that, without velocity information, the non-Magellanic HVCs would be hard to distinguish from the flared, warped disk.
}
\label{fig:other}
\end{figure}

Many of the halo HI features that can clearly be disentangled from the disk of the spiral are potentially linked to satellite accretion or interaction with a companion - e.g., M31 \citep{westmeier05, ibata07}, NGC 891 \citep{oosterloo07, mouhcine10}, M33 \citep{putman09, grossi08}, NGC 5055 \citep{battaglia06, martinez10}, NGC 253 \citep{boomsma05, beck82}, NGC 2442 \citep{ryder01}; and see \cite{hibbard01} and \cite{sancisi08} for other examples.   As can be seen from the bottom panels in Figure~\ref{fig:other}, this would also be the case for the Milky Way with only the Magellanic System HVCs being distinct if it were observed externally.  Future deep stellar observations may link additional HI halo features to satellite accretion; however, this is not possible when the HI feature overlaps with the disk in projection - e.g., NGC 2403 \citep{fraternali02, fraternali01}; NGC 4395 \citep{heald07}.

The high resolution observations of galaxies designed to detect halo clouds often have a HI column density sensitivity of  $\sim10^{19}$ cm$^{-2}$.  This is only the typical peak column density of a Milky Way HVC and therefore may be missing a significant amount of halo gas.
The spiral galaxy that has been mapped to the deepest levels (N$_{HI} < 10^{18}$ cm$^{-2}$) at a resolution of a few kpc, and is most similar to the Milky Way, is M31 \citep{thilker04, westmeier05}.   NGC~891 has also been mapped to $10^{18}$ cm$^{-2}$ levels at this resolution, but this galaxy has a much higher SFR than the Milky Way \citep{oosterloo07}.  Both of these galaxies are shown in the top panels of Figure~\ref{fig:other}.  The majority of their halo gas is connected to the HI disk (like most other galaxies), and/or could be associated with a satellite interaction.  The calculated HI mass in M31's halo is the same as that calculated for the Milky Way without the Magellanic System ($3-4~\times 10^7$ \Msun; see \S\ref{sec:phys}), and presumably these clouds also have a substantial ionized component similar to the Milky Way's HVCs \citep{fittingoff09}.   NGC~891 shows a larger amount of halo and disk-halo gas, which is consistent with its higher star formation rate \citep{oosterloo07}.  A diffuse halo medium can be inferred to exist from the head-tail nature of several of the M31 HVCs \citep{westmeier05}.   The M31 HVCs also have similar linewidths to Galactic clouds.

\subsection{Warm Ionized and Warm-Hot Halo Gas}
\label{sec:otherwarm}

Similar to the Milky Way, the warm gas in the halos of other spiral galaxies is detected in emission and absorption.   The emission line observations primarily detect gas close to the disk of the spiral \citep[within $\sim$5 kpc;][]{hoopes99,collins01}, and show that the extent of the extraplanar H$\alpha$ emission scales with the galaxy's SFR \citep[e.g.,][]{rand96, rossa03}.  Investigations of the motion of the ionized gas find a similar rotation lag to the HI with increasing z-height \citep{rand00,heald06}.  \cite{heald07} find a typical vertical gradient of magnitude 15-25~\kms~per scale height.  The gas increases in temperature with z-height \citep{collins01,rand11}, as also found for the Milky Way.  See \cite{haffner09} for a further review of the warm ionized gas of spiral galaxies detected in emission.

There are abundant absorption line results that are presumed to trace the warm gas in galaxy halos based on their observed kinematic and spatial proximity.  Since low column density absorbers (N$_{HI}$ \lessapprox $10^{15}$ cm$^{-2}$) are also detected along large scale cosmic filaments, the association with a galaxy halo is not certain \citep{bowen96, putman06,penton02}.  Within $\sim300$ kpc of a galaxy, Ly$\alpha$ absorber results show a covering fraction close to 100\% and total gas mass estimates are $10^{9-10}$ \Msun~\citep{bowen02,wakker09, prochaska11, chen00}.  
 For the higher column density Ly$\alpha$ absorbers (and Mg II absorbers that are generally thought to trace gas with N$_{HI} > 10^{16}$ cm$^{-2}$), a relationship to the halos of galaxies is relatively clear as the detections are almost solely in the vicinity of galaxies \citep{bowen02, penton02, chen10b}. 
There are a limited number of these high column density absorbers at $z=0$ around spiral galaxies.
 Studies of Mg II absorbers at $z\sim0.1-1$ suggest the covering fraction is $>50$\% within 100 kpc of the galaxy's center \citep{chen10b,kacprzak08}; although the covering fraction is expected to be higher at higher redshifts \citep{fernandez12}.

Studies of the warm-hot component ($\sim10^{5-6}$~K) are largely limited to O~VI absorbers at $z=0$ \citep{prochaska11,wakker09}, but will soon be extended to include C~IV \citep{green12}.  Broad Ly$\alpha$ absorbers have also been claimed to be warm-hot gas \citep{bowen02, savage11, danforth10}.  The low redshift O~VI absorbers have significant covering fractions of $\sim70-80$\% within 200-400 kpc of galaxies \citep{prochaska11, wakker09}.   This seems to be consistent with the extensive work at higher redshifts that measure covering fractions close to unity for C~IV ($z\sim0.5$) and O~VI ($z>0.1$) within 100 kpc of star forming galaxies \citep{chen01, tumlinson11, chen09}.   Star forming galaxies are more likely to have both warm and warm-hot gas in their halos \citep{tumlinson11,chen10}.
\cite{tumlinson11} estimate a mass of circumgalactic gas of 2$\times 10^9$ \Msun (assuming solar metallicity) out to 150 kpc from the abundant O~VI detections near galaxies at redshifts z = 0.1-0.36.  
This is similar to the mass derived from the Ly$\alpha$ absorber detections in galaxy halos noted above, but approximately an order of magnitude higher than the mass estimates for ionized gas in the Milky Way halo (\S\ref{sec:warm}).  This may be due to several reasons:  1) extragalactic absorption line observations may also capture gas in cosmic filaments near the galaxy, but beyond the virial radius, 2) the assumptions used to calculate total gas mass from extragalactic absorption line results lead to an over-estimate, 3) a range of galaxy types are used for extragalactic measurements, or 4) the Milky Way warm and warm-hot halo gas is less massive or has not yet been fully observed.

\subsection{Hot Halo Gas}
\label{sec:otherhot}

The x-ray detected hot gas in the halos of Milky Way-like spiral galaxies always extends from the galaxy's disk and is within a 10 kpc radius \citep{li08, li06, wang03, immler03,otte03,wang01, bregman97}.   The scale heights are calculated to be 1-2 kpc for many spiral galaxies, and the x-ray features are similar to the extraplanar features detected in H$\alpha$ emission \citep{strickland04,bregman97, tyler04}.  There are numerous limits on the amount of hot halo gas at large radii from non-detections \citep[e.g.][]{yao10, anderson10, rasmussen09}.  As with the Milky Way, most researchers argue that the hot halo medium cannot account for a large percentage of a spiral galaxy's baryons.  If galaxies substantially more massive than the Milky Way are considered, the hot halo gas extends to larger radii \citep[e.g., $\sim$50 kpc for NGC 1961 with M$_{\rm tot} > 10^{13}$ \Msun;][]{anderson11}.
Unlike the Milky Way, there is limited indirect evidence for an extended, diffuse, hot halo medium around other spiral galaxies (although see \S\ref{sec:otherwarm} for the warm-hot gas).  One exception is M31, where the head-tail HI clouds out to $\sim30$ kpc from the disk are likely to originate from the movement of cold clouds through a diffuse, hot halo medium \citep{westmeier05}.

\subsection{Dust, Metals and Molecules}

Direct detections of dust are primarily at the disk-halo interface of spiral galaxies \cite[e.g.,][]{keppel91}. A survey of nearby edge-on galaxies found dense dust ($A_V \sim 1$) at least 2 kpc above the plane, and the dust is correlated with the presence of extraplanar diffuse ionized gas (DIG) \citep{howk99}.  \cite{menard10} have made a convincing detection of dust in extended galaxy halos using the reddening of SDSS background quasars.   They found substantial reddening 
20 kpc from the galaxies, with detectable reddening all the way out to 10 Mpc \citep[see also][]{zaritsky94}.
Under an assumption of SMC-like dust, they claim that $L_*$ galaxies have similar amounts of dust in their halos and disks, consistent with observations of reddening in Mg II absorbers at intermediate redshift \citep{menard12}.

Metals are also present in spiral galaxy halos, as directly evident from their detection with absorption lines (\S\ref{sec:otherwarm}).   Though ionization conditions are difficult to assess, \cite{tumlinson11} estimate there is $\sim10^7$ \Msun~in oxygen within 150 kpc of star forming galaxies at $z=0.1-0.36$ using O VI measurements.  The abundant detections indicate enriched gas has been fed into halos for some time.   Sightlines for absorption line experiments that pass through extragalactic HI halo features are rare, but in the cases where star formation has occurred in the gas, metallicities can be measured with the HII regions.   Most of this type of halo gas is related to an interaction and the metallicity of the halo HI features is similar to the outer region of the galaxy \citep{werk11}.
These interacting systems with halo star formation are also the only ones for which molecular gas has been detected or inferred to exist \citep[e.g.][]{lee02}.

\section{Origins}
\label{sec:origins}

Spiral galaxy halo gas is likely to originate from multiple sources, including from the intergalactic medium, satellite galaxies, and the disk of the galaxy itself.    
This section discusses these origins in the context of the observational results presented in \S\ref{sec:ghalo} and \S\ref{sec:other}.   Viable origin scenarios should be consistent with the following key observational features.
\begin{itemize}
\item{Large scale flows of multiphase gas are prevalent in galaxy halos.  This is especially evident from the spatial and kinematic link between the cold, warm, and warm-hot gas in the Milky Way halo (see Figures~\ref{fig:assoc} and \ref{fig:assoc2}).}
\item{The densest halo gas is closest to the disk in all phases.   Exceptions (and the largest cold halo features) are largely linked to satellite accretion (see Figures~\ref{fig:3d} and ~\ref{fig:other}).}
\item{Gas at the disk-halo interface shows a clear temperature gradient (cold to hot), and kinematic gradient (decreasing rotation) with increasing z-height (see Figure~\ref{fig:diskhalo}).   This gas is also more enriched than gas further out in the halo and more extensive in galaxies with higher SFRs.}
\item{Diffuse gas extends to radii $>100$ kpc in spiral galaxy halos.  This is evident for the Milky Way from the properties of clouds associated with the Magellanic System and stripped dwarf galaxies, and for other galaxies from absorption line results that show large covering fractions.}
\item{For the Milky Way, the mass in warm and cold halo gas is $<10^{9}$ \Msun, and, though largely model-dependent, the hot gas mass is $<10^{10}$ \Msun.
These measurements are consistent with the HI and x-ray observations for other low redshift spiral galaxies of similar mass.}
\item{The halo gas appears to be bound to the dark matter halos of the spiral galaxies.   This gas is also found to be enriched with metals at some level (i.e., non-primordial).}
\end{itemize}

\begin{figure}
\includegraphics[trim=130 115 50 100, clip=true, scale=1]{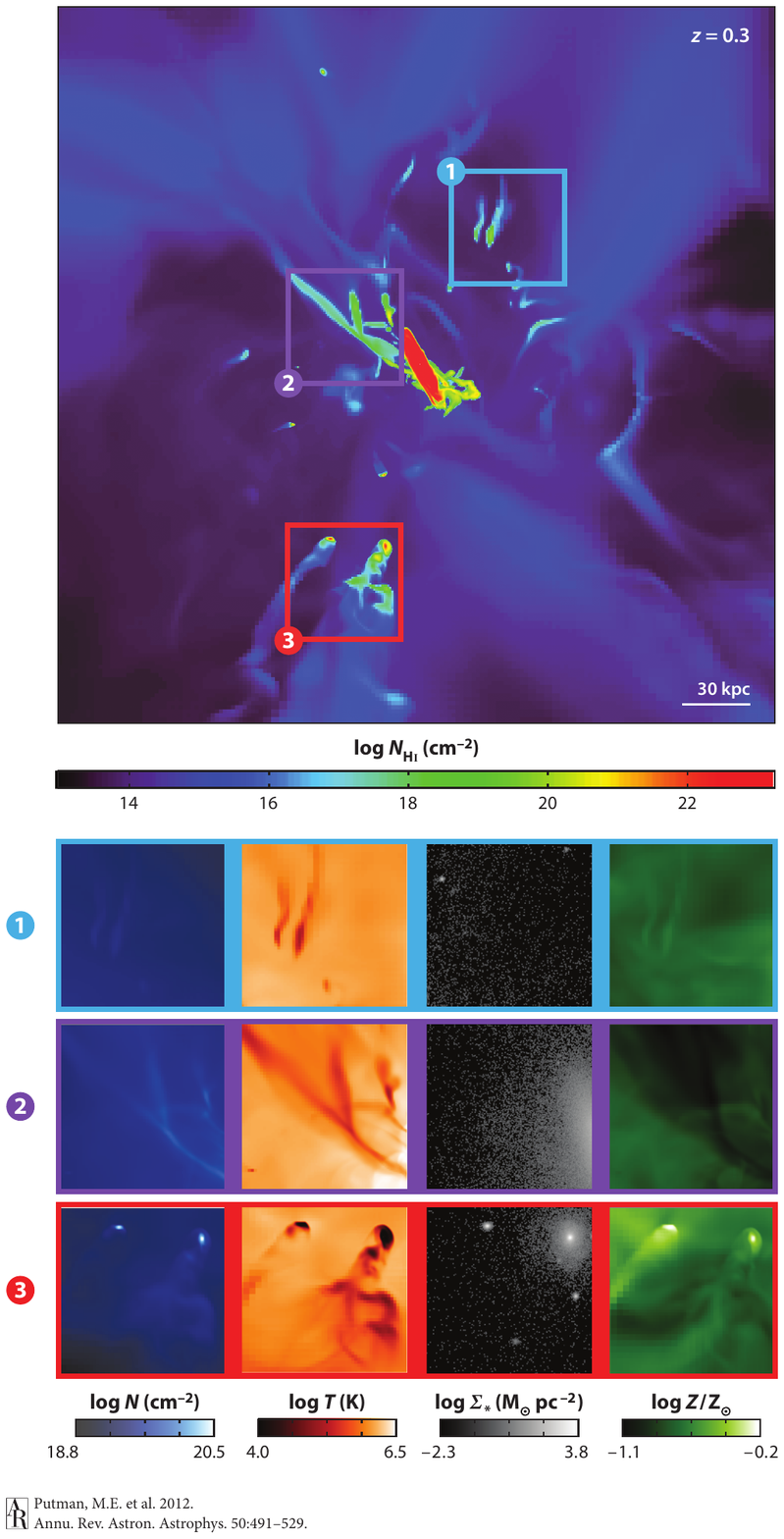}
\caption{\scriptsize A cosmological AMR simulation of a Milky Way galaxy at z=0.3 \citep{joung12b}.  We chose $t = 10.3$ Gyr as a relatively low redshift time slice, as it clearly shows examples of three different origins of neutral gas in the halo as indicated by the numbered boxes in the top panel.  Box 1 represents the interaction between outflowing gas and accreting halo gas, box 2 shows a filamentary cool flow feeding the disk, and box 3 represents the accretion of three satellites.   The cut-outs show the total gas density (N), temperature (T), stellar density ($\Sigma_{*}$) and metallicity (Z).  The features become less distinct in the total gas density plots due to the large amount of multiphase gas in the halo.}
\label{fig:sim}
\end{figure}

\begin{figure}
\includegraphics[trim=120 100 50 110, clip=true, scale=1]{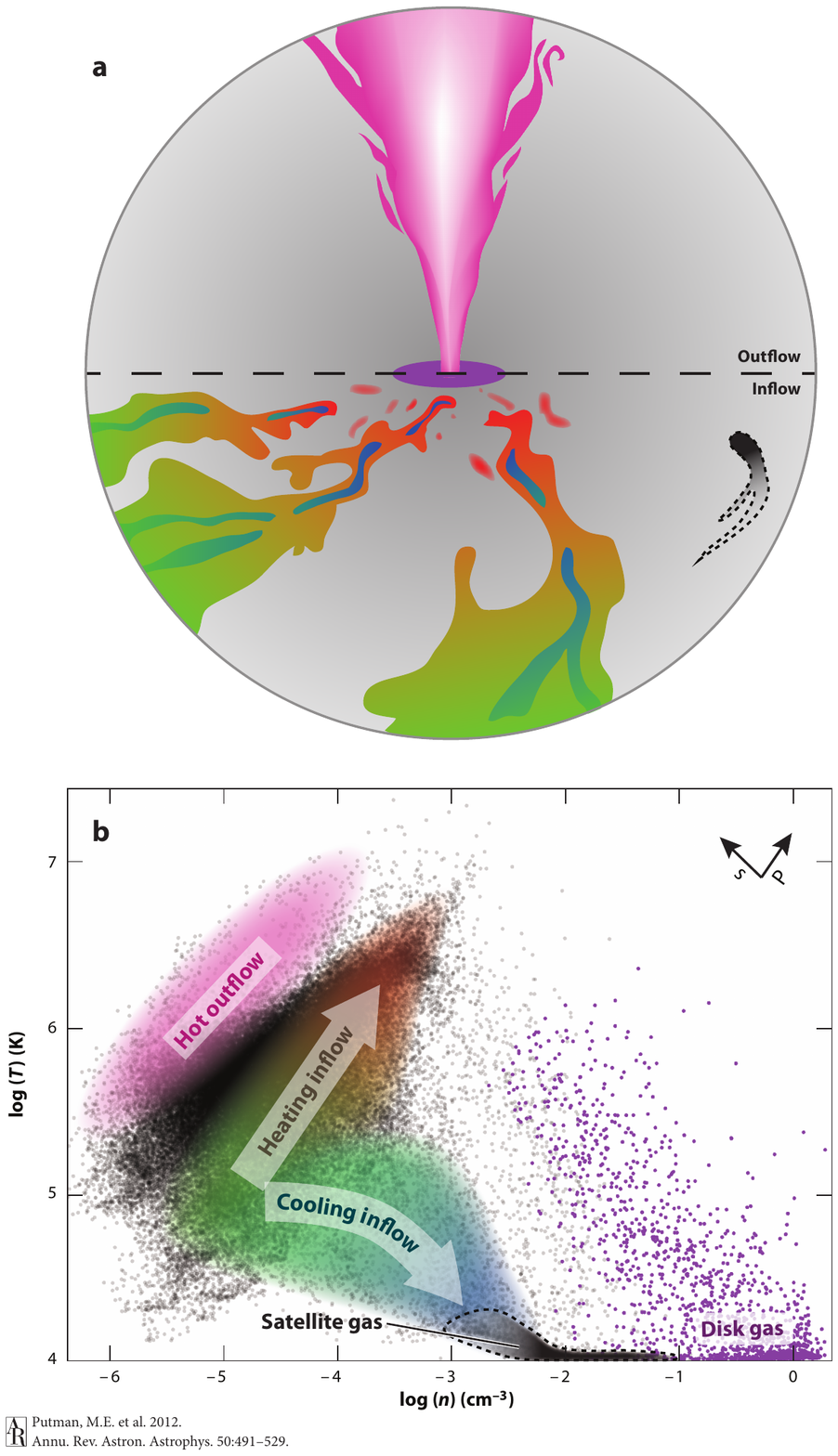}
\caption{\scriptsize Top:  A diagram based on the MW simulation at z=0 to a radius of 100 kpc where the top hemisphere shows the outflow due to SN-driven winds (pink) and the bottom shows the inflow of cooling (green to blue) and heating (green to red) filamentary flows and a satellite (black).    Bottom:  The density-temperature diagram for cells in the MW simulation at z=0 with color coding corresponding to the top panel and main features noted.   The directions of increasing pressure and entropy are shown with the arrows in the upper right.  The cells were randomly selected from the simulation with the probability of selection proportional to gas density.}
\label{fig:simcart}
\end{figure}

Throughout this section we refer to and make connections to a cosmological grid simulation of a Milky Way-mass galaxy at $z=0$ \citep[referred to as MW simulation;][]{joung12b,fernandez12}.  We use this simulation not as the final word on the state of the Milky Way halo, but rather as a guide to our understanding of the important physical processes at play.  The simulation, performed with an adaptive mesh refinement (AMR) code, Enzo 
\citep{bryan97,oshea04}, has sufficiently high spatial (136--272 pc comoving or better at all times) and mass  ($\sim10^5$ \Msun) resolutions to allow us to study and track the spatial and kinematic distribution of the multiphase gas in the halo in detail.  The simulation includes metallicity-dependent cooling, metagalactic UV background, shielding of UV radiation by neutral hydrogen, photoelectric heating for the diffuse gas, star formation and stellar feedback (but not AGN feedback).  Figure~\ref{fig:sim} displays a snapshot of the MW simulation in HI at $z=0.3$ and zooms in on halo gas origins relevant to each of the below sections.  
Figure \ref{fig:simcart} shows a cartoon based on the simulation that is split to show the main accretion and feedback mechanisms distinctly.   It also depicts the density-temperature diagram of simulation cells at z=0.

\subsection{Accretion from the IGM}
\label{sec:coldflow}

Under $\Lambda$CDM, the large majority of the baryons enter the virial radius of a galactic halo in gaseous form along cosmic filaments.  This is in contrast to the classical picture that posited that the incoming gas would spherically collapse and be shock-heated close to the virial temperature of pressure-supported gas within a galaxy's halo \citep{white78, white91}.
Analytic arguments \citep{birnboim03, dekel06, binney77} and SPH simulations have shown that some of the gas will never be shock-heated to the virial temperature of the halo and be accreted via a ``cold mode" \citep{keres05, brooks09}.  
This mode is found to dominate in galaxies at high redshift ($z$ \gtapprox~ 2), as well as in present day low-mass galaxies ($M_{\rm halo}$ \lessapprox ~$10^{12}$ \Msun).  
We note that this mode is referred to as cold, but in reality the gas is at temperatures \lessapprox~$10^{5.5}$ K, so much of it is not cold under our observational definition in \S\ref{sec:ghalo}. 

The distinction between hot and cold mode accretion is based on the maximum temperature that a given gas particle attains along its trajectory as it enters the virial radius.   At $z\sim0$, for galaxies with masses similar to the Milky Way, a large fraction of the accreted gas is hot mode and does not connect directly to the disk.  The hot mode gas is heated either by one strong shock close to the virial radius or via a series of weak shocks as the kinetic energy of the outer portions of the filaments is converted to thermal energy due to compression from the existing halo gas \citep{keres05, joung12b}. 
In the MW simulation, $\sim70$\% of the mass inflow at the virial radius is consistent with coming in along filaments.
The cold mode of the accretion at $z\sim$0 is found in the core regions of the compressed filaments at temperatures \lessapprox~$10^{5.5}$ K.   This gas stays cold because it is shielded from interacting with the existing halo medium by the hot mode gas that surrounds it in the filamentary flow \citep{joung12b}.  The accretion of the hot and cold mode gas along filamentary flows is depicted in the top panel of Figure~\ref{fig:simcart} and box 2 of Figure~\ref{fig:sim}.

The properties of the filamentary flows can be examined further in the MW simulation.  In the density-temperature diagram of Figure \ref{fig:simcart}, the filamentary flows generally have intermediate specific entropy between that of the disk gas and the outflows.  The flows come in at temperatures of $\sim10^{4.5-5.5}$~K and the outer portions are heated toward $\sim10^6$~K and the inner portions cool toward $\sim10^4$~K.
The gas at $< 10^{5.5}$~K has a relatively high neutral fraction (generally $f_{neut} > 10^{-4}$), and extends from radii of approximately 25 - 200 kpc.  Some of the cold cores of the filaments have HI densities comparable to those of gas stripped from simulated satellites, but they generally have lower metallicities ($Z/Z_{\odot}$ \lessapprox~0.2; see Figure~\ref{fig:sim}).  
Details on the HI properties of the halo gas in the MW simulation are outlined in \cite{fernandez12}.

In terms of observations, the filamentary flows could be represented by the diffuse warm and warm-hot gas detected with absorption lines near the velocity of the galaxy (\S\ref{sec:otherwarm}); the majority of these filaments are not detectable in HI with our current sensitivities until close to the disk (see below).  The covering fraction of the observed warm halo gas approaches 100\%, and this is consistent with the covering fraction in the MW simulation when low enough column density gas is selected \citep[$< 5 \times 10^{13}$ cm$^{-2}$;][]{fernandez12}.  
The gas that is heated to temperatures $> 10^{5.5}$ K largely becomes part of the general hot halo medium, and may be related to the fact that we see evidence for a diffuse hot halo medium out to large radii (however see also \S\ref{sec:feedback}).  The MW simulation finds $2 \times 10^{10}$ \Msun~ in hot gas within the virial radius and this is consistent with the simulation of \cite{sommer06}.  This is greater than the amount found by observations thus far. 

For the gas in the filamentary flows to feed the star-forming disk, the gas needs to cool further; the gas potentially forms clouds within the halo \citep{maller04, connors06, kaufmann06, kaufmann09, lin00}.  The cold mode gas contains density enhancements that can cool as long as the cooling time is shorter than the disruption time \citep{joung12, binney09}.  Linear fluctuations were not able to cool in these detailed studies and cooling without significant density enhancements is also not seen in the MW simulation (Figures~\ref{fig:sim} \& \ref{fig:simcart}).   The gas that does cool may not survive the trip to the disk without disrupting further (see \S\ref{sec:survive}).  It remains somewhat unclear how halo gas accreted in hot mode can feed the disk.  Possibly closer to the disk, where it is denser, the hot gas may mix with feedback material and become part of the disk-halo interface (\S\ref{sec:diskhalofeed}).  In any case, in the MW simulation the gas is only observable as HI clouds within $\sim50$ kpc, and this is consistent with observations and previous simulations \citep{sommer06,kaufmann06, keres09}.   The mass in observed warm and cold clouds is $10^{8-9}$ \Msun~ in both the observations and simulations when the same type of gas is selected \citep{peek08,fernandez12}.

\subsection{Feedback}
\label{sec:feedback}

After baryons collect at the centers of galaxies and form stars, the stars
and central black holes act back on their environments and can affect the gaseous galactic halos.  These ``feedback" processes include radiation
from young stars that heat and ionize the halo gas \citep{blandhawthorn99, bland02, bland01}, mechanical energy from
supernova explosions \citep[e.g.,][]{maclow04,veilleux05} and AGNs \citep{antonuccio10, dimatteo05, dimatteo08}, as well as
momentum-driven winds due to radiation pressure from central sources on
dust grains \citep{murray05} and a galactic fountain as enriched hot gas rises up from the disk \citep[e.g.,][]{shapiro76, joung06}.  We will discuss each of these mechanisms in turn.  Feedback is required in cosmological simulations to reproduce the fundamentals of observed galaxies; however, despite its importance, it is described by phenomenological prescriptions partially due to our lack of understanding of the dominant process(es) \citep{oppenheimer06, springel03, stinson06, dalla08}.
Halo gas is in many ways a more direct consequence of disk feedback than the stellar observables of galaxies and is therefore an important tool for understanding these processes. 

Radiation escaping from a galaxy's disk ionizes and heats halo gas.  H$\alpha$ observations of cold Galactic
halo clouds have been used to estimate that 6\% of the Milky Way's ionizing photons escape normal to the disk \citep[f$_{\rm esc}$ = 1-2\% when averaged over the solid angle;][and see \S\ref{sec:warm} and \S\ref{sec:phys}]{blandhawthorn99}.  
For other spiral galaxies, there are mainly upper limits \citep[e.g.,][]{tumlinson99, deharveng01}, 
and the measurements are consistent with escape fractions of less than a few percent as long as starbursts are excluded.   As expected, if f$_{\rm esc}$ is estimated with a galaxy's diffuse ionized gas it is highly dependent on the clumpiness of the galaxy's ISM \citep{clarke02, zurita02, fernandez11}.   
Radiation escaping from galaxy disks is only starting to be included in cosmological galaxy formation simulations.
Constraints on the escape fraction impact both halo gas and the ionization of the IGM as a whole, in particular at higher redshift \citep[e.g.,][]{ciardi03, gnedin00}.

Mechanical feedback mechanisms, such as SN-driven or past AGN-driven outflows, are the most relevant in the context of the origin of a spiral galaxy's halo gas.   For the Milky Way, a diffuse hot halo medium is evident out to the distance of the Magellanic System ($\sim 50-100$~kpc), and this indicates either relatively strong Galactic winds and/or gas left from hot mode accretion is present.   The detected x-ray and gamma-ray 'bubbles' are also consistent with the Milky Way having feedback from a central engine in the past 10-15 Myr \citep{su10}.  For other galaxies, outflows are consistent with the enriched warm-hot gas at distances on the order of the galaxy's virial radius (\S\ref{sec:otherwarm}).   Finally, the baryon fraction within the virial radius of Milky Way-like spirals is substantially smaller than the cosmic mean ($\sim0.4$ for the Milky Way), and this is attributed to powerful outflow mechanisms \citep{bregman07, mcgaugh10}, as is the ability to reproduce the basic properties of a galaxy in cosmological simulations \citep{benson03, mccarthy10, mccarthy11}. 

Some of the material from large scale galactic winds is likely to escape the galaxy's halo (potentially explaining the lack of baryons), and a fraction is also thought to feed the hot halo.   
In Figure~\ref{fig:simcart} we show a cartoon based on the MW simulation that depicts outflow due to SN-driven winds.    This wind extends from velocities of 200\kms~to over 1000\kms.  The fastest gas escapes, while the lower velocity gas feeds the hot halo high entropy, enriched material.  Observationally, the lower velocity material may be some of the low density absorption line systems in the Milky Way and other galaxy halos (\S\ref{sec:warm} \& \S\ref{sec:otherwarm}); in particular, the gas at high positive velocity in the general direction of the Galactic Poles shown in Figure~\ref{fig:assoc2}.  The SN-wind gas at higher velocities is too hot and low density to be observable.   Figure~\ref{fig:simcart} also shows that the MW simulation confirms previous simulation expectations that supernova feedback can be thought of as injecting high entropy gas into the system.  In particular, once launched, the specific entropy of this enriched gas remains unchanged for a large range of the galactocentric radius, suggesting adiabatic expansion.  

Momentum driven feedback is also commonly discussed as a method of launching material from the disk.
This mechanism is likely to be relevant only in galaxies with powerful starbursts or AGNs \citep{oppenheimer06, murray07} that may have operated in
the more violent, formative years of present-day spiral galaxies like the Milky Way \citep[$z \sim 2-3$; e.g.,][]{kim11}. 
In addition, this mechanism may have difficulty launching material as, even if one assumes the maximal efficiency of converting mass to energy 
and that all the available momentum is used to drive outward motion of the surrounding gas, the outflow velocity would be too low to eject gas for reasonable values of the mass loading factor.    Momentum driven winds have also been invoked as a method to launch substantial amounts of dust into galaxy halos \citep{aguirre01}; however, energy-driven mechanisms may be equally effective through the coupling of dust and gas.    Dust observations thus far could be consistent with either mechanism.

A galactic fountain is distinct from the powerful galactic winds discussed above in that it is a disk-wide phenomenon in which gas is driven upward by the cumulative effect of supernova explosions and adiabatic
expansion, and subsequently falls back down due to radiative cooling \citep{shapiro76, bregman80}.
The galactic fountain phenomena is most likely to affect the detailed structure and dynamics of
gas at the disk-halo interface.  
Kiloparsec-scale ISM simulations in vertically elongated boxes
\citep{deavillez00, deavillez05, joung06, booth07, hill12} report ejecta cooling and condensing out at a few kpc heights having
morphology and kinematics similar to those of observed IVCs (\S\ref{sec:diskhalo}).  The near-solar metallicities and dustiness of many IVCs are also consistent with a disk origin.  The temperature gradient of the disk-halo interface gas is expected with hot gas from the SN rising to higher z-heights, and the kinematic lag suggests a combination of this type of feedback and fueling (see \S\ref{sec:diskhalofeed}).   The large population of small disk-halo clouds may also be linked to star formation processes in the disk \citep{ford10}. The cartoon shown in Figure~\ref{fig:diskhalo} summarizes some of the observational findings for gas at the disk-halo interface and has many similarities to the fountain simulations.

Feedback material mixes and interacts with the existing halo gas.  
In box 1 of Figure~\ref{fig:sim}, we show an example in the MW simulation of neutral halo clouds that have originated from the interaction of a supernova-driven wind and incoming cooler flows.  The gas has condensed at distances of 50-100 kpc from the disk and has metallicities around 0.3 $Z_{\odot}$, which is greater than many of the other simulated HI halo features, and lower than that of the wind itself.
The HI clouds fragment further and impact the disk $\sim$300 Myr later than the timestep shown.  
The higher metallicities of some regions in HVC Complex C may be representative of this mixing process closer to the disk.   HVC interaction with feedback material may also be evident in the H$\alpha$ brightness of some complexes and the mixture of photo- and collisional ionization required to model the line ratios measured for halo gas.

\subsection{Satellite Accretion}
\label{sec:sat}

Satellites lose their gas as they move through a spiral galaxy's halo \citep[e.g.,][]{putman03,grcevich09}.   The gas is primarily ram pressure stripped from the dwarf in models, although tidal forces also play a role \citep{mayer07, mayer06}.  
 $\Lambda$CDM simulations find 40\% of L$_*$ galaxies host a 0.1 L$_*$ satellite within its virial radius and this is consistent with results from SDSS \citep{tollerud11,liu11}.   This percentage drops significantly at smaller radii (e.g., 12\% within 75 kpc) and also when only blue and star forming satellites like the LMC are considered \citep[8.2\% within 100 kpc;][]{tollerud11, james11}.   We know from the Local Group that the vast majority of the satellites have $< 0.1$ L$_*$, however the amount of gas they bring is substantially smaller ($\sim10^{5-8}$ \Msun~each).  For instance in the Local Group, the Magellanic System contains $\sim30-50$\% of the HI mass of the Milky Way, and M33 will eventually provide $\sim25$\% of M31's HI mass \citep{putman03,putman09}.   The lower mass satellites within the virial radius of the Milky Way will only have provided $\sim1$\%, or if we include the large number of yet unobserved satellites predicted by $\Lambda$CDM, $\sim10$\% \citep{grcevich09}.

The cumulative effect of the accretion of satellites will leave behind a large amount of gas that will eventually feed the disk.   Since most galaxies are likely to be stripped at radii $>$ 20 kpc  \citep{grcevich09,mayer06,nichols11}, much of the cold gas will be integrated into the diffuse galactic halo before eventually recooling at the disk (see \S\ref{sec:survive}).  This is consistent with multiphase flows being found throughout galaxy halos and helps to sync the angular momentum distribution of the satellite gas with the spiral's halo gas.  This integration of satellite material with the halo medium can be seen in box 3 of Figure~\ref{fig:sim} which depicts the accretion of a triple system of dwarf galaxies.
The stripped gas shows a temperature gradient as it integrates into the hot halo material, and the column density and metallicity of the gas also gradual decreases (see also Figure~\ref{fig:simcart}).  The cold and warm satellite gas has on average higher radial velocities than other features in the halo.  This material becomes untraceable several hundred Myrs after being stripped from the satellite core, but may leave behind overdensities that help seed the cool filaments closer to the disk (see \S \ref{sec:survive} and \S\ref{sec:diskhalofeed}).  

The system shown in Figure~\ref{fig:sim} has many similarities to what is observed for the Magellanic System of the Milky Way, which has given us a detailed look at the relationship between gaseous satellites and a spiral galaxy's halo gas (see Figures~\ref{fig:assoc} - \ref{fig:3d} and \ref{fig:other}).   The Magellanic System shows a column density gradient along its length and multiphase gas surrounding it, particularly at the tail where it is being integrated into the hot halo medium.  This system can also account for the highest negative and positive velocity halo gas in the Milky Way halo.  The interaction of the satellite gas with an existing halo medium, and the high covering fraction of the warm and warm-hot gas does make it relatively clear that mechanisms besides satellite accretion are also feeding the halo.  In addition, for the Milky Way, most of the HVCs (besides those associated with the Magellanic System) do not show any clear link to satellites or stellar streams.  There is a large body of literature describing the Magellanic System and it is thought to be the result of a combination of a tidal interaction between the two satellites and their subsequent interaction with the Milky Way \citep{connors06b, mastropietro05, besla10, gardiner99, diaz11, nidever08}.    

In general if the HI gas observed in galaxy halos is directly from a satellite, it is likely to have been relatively recently stripped from the satellite.  This is consistent with the link between large HI features in spiral galaxy halos and a relatively recent interaction event (see \S \ref{sec:otherhi}).   Satellite HI features are also likely to have higher metallicities than gas that originates from the IGM \citep[e.g.,][]{fernandez12}, although given the likely mixing of halo gas components, deep maps of the stellar components of galaxy halos may remain the best method to identify a satellite origin for the gas \citep[e.g.,][]{martinez10}.  The Milky Way has a large number of HVCs at z-heights \lessapprox~10 kpc that do not seem to be related to stellar halo features.   These clouds may be a combination of satellite gas and halo gas of other origins that is recooling close to the disk.

\section{Halo Cloud Survival}
\label{sec:survive}

Cold halo clouds moving through hot diffuse galaxy halos are unlikely to survive for periods greater than a few hundred Myr unless a strong support mechanism is invoked (see below).   These short lifetimes are a problem unless the clouds are continually regenerated on short timescales.  In addition, some HVCs that are not clearly linked to satellite accretion have velocities that cannot be reached through gravitational acceleration within several hundred Myr \citep{benjamin97,peek07}.   The exact lifetime of a cloud depends on several factors such as cloud density, halo density, and velocity, but the total mass of the cloud seems to be one of the largest factors increasing their lifetimes \citep{heitsch09,kwak11}.  This is consistent with simulations that show the halo clouds are destroyed primarily via the Kelvin-Helmholtz (KH) instability.  The characteristic growth time for the KH instability is $t_{KH} \propto \chi^{1/2} R_{cl} / v_{rel}$, where $\chi$ is the density contrast between the cloud and the external medium, $R_{cl}$ is a characteristic length of the cloud, and $v_{rel}$ is the relative speed between the two.  The KH instability leads to gas being ablated from the cloud edges and forming a tail of lower column density material.  Observationally this is evident in the head-tail clouds found throughout the Galactic halo \citep{putman11,bruens00,bekhti06, westmeier05b}, and potentially some of the detailed structure found along the edges of larger clouds \citep{peek07, winkel11} as shown in Figure~\ref{fig:ht}. 
O~VI observations are also consistent with the interaction of halo clouds with the hot diffuse halo medium \citep{sembach03, kwak11, fox06, collins07}.

Possible methods of supporting the clouds to significantly extend their lifetimes include the presence of dark matter \citep{nichols09, braun00}, magnetic fields \citep{mcclure10}, and/or shielding by an extended diffuse gaseous component (e.g., the extended medium surrounding the Magellanic Stream).  The presence of dark matter extends a gas cloud's  lifetime to timescales on the order of a Gyr \citep{quilis01}; however despite the advantage of this preservation there is no strong evidence for dark matter in HVCs.  Rather, even the small compact clouds are associated with large gaseous complexes \citep[Figure~\ref{fig:assoc};][]{saul12,putman11}, and it would be difficult for gas to survive in such small dark matter halos through reionization \citep[e.g.,][]{ricotti08}.   A magnetic field may help to preserve the clouds \citep{konz02, santillan04, kwak11}, however most of the simulations done thus far are in two dimensions and some simulations indicate a strong magnetic field can aid in the disruption \citep{stone07}.

Dynamical shielding by outer envelope layers may be the most likely mechanism of extending the lifetimes of HI clouds.  As shown in Figures~\ref{fig:sim} and \ref{fig:simcart}, the MW simulation finds cold neutral clouds are embedded within larger, warm, low density structures and Figure~\ref{fig:assoc} shows that this is also observed. 
In this configuration, the HI clouds experience smaller relative velocities with respect to their immediate surroundings, which increases $t_{KH}$.  Head-tail structures or features indicative of shearing may be less common in the HI component of those clouds that are dynamically shielded by an outer, warmer layer.  This may explain why clouds associated with the Leading Arm of the Magellanic System do not have detectable warm gas and also show abundant head-tail structures, while the tail of the Magellanic System has abundant warm diffuse gas and limited head-tail structures \citep[e.g.,][]{putman11}.

For the Milky Way, since many of the HI halo clouds are at distances of $\sim10$ kpc (or greater in the case of the clouds associated with the Magellanic System), and since the individual clouds within the complexes usually have masses $<10^5$~\Msun, the clouds are unlikely to make the trip through the Galactic halo unless significantly shielded by an outer, warmer component.  For other galaxies, the detected large HI clouds are also likely to be collections of smaller clouds.  The disrupted HI clouds will become part of the multi-phase halo medium, and any remaining over-densities will sink toward the Galactic disk.  Some of the warm and warm-hot gas detected with absorption line systems may be the leftover over-densities from disrupted HI clouds.   Whether these over-densities are able to re-cool depends on the magnitude of the over-density, and, if they cool, they may disrupt again before impacting the disk  \citep{joung12, vietri97}.   The infall of these clouds can dissipate energy and help to heat the extended hot halo medium \citep{murray04}.  According to multi-wavelength observations, gaseous halos are complex multiphase places, which is consistent with an ongoing cycle of halo cloud destruction and cooling.

\section{Feeding Galaxy Disks}
\label{sec:feeding}
The amount, metallicity, and ionization state of the halo gas has consequences on a galaxy's resulting star formation.   
Star formation rates of spiral galaxies at low redshift are approximately 0.5 - 5 \Msun$/$yr, and Galactic chemical evolution models find that at least half this amount needs to be continuously accreted, low metallicity fuel \citep{chiappini09, chiappini01}.   Without accretion, the SFRs indicate the star formation fuel currently in the disk will last for another few Gyrs at most.
In this section, potential sources to maintain a galaxy's star formation over time are discussed.    This includes a calculation of the accretion rate for Milky Way halo gas, a discussion of gas at the disk-halo interface, and a brief mention of some often neglected disk sources.

\subsection{Halo Sources}

\label{sec:accrete}

The mass inflow rate to the Galactic disk is a crucial parameter to derive from the physical properties of the HVC complexes outlined in \S\ref{sec:phys}.    Many studies estimate the HVC contribution to the accretion rate using some form of the equation,
\begin{equation}
\dot{M} = \sum \frac{M_i v_{i}}{ D_i}
\end{equation}
where $M_i$ is the mass of a given complex, $v_i$ is the observed radial velocity and $D_i$ is the distance \citep{thom08, wakker07}. The largest difficulty with this method is that the observed radial velocity of a HVC complex is the sum of the velocity of the complex toward the disk (or more generally a poloidal component), and the azimuthal speed of the cloud around the disk (toroidal component), projected along the line of sight. This observed velocity may be very different from the true radial velocity of the gas toward the Galaxy, and the cloud may be accelerated in the future by MW gravity and/or decelerated by ram pressure.  Additionally, the distance to the gas is not the same as the distance from the cloud to where it will eventually impact the disk.

Now that most of the complexes have direct distance constraints, we can construct a very simple model of the HI accretion process that takes into account the bulk of the accretion onto the Galaxy from known HVCs in the next $\sim$ Gyr. We only consider the 13 complexes with relatively well-known distances, excluding the Magellanic Stream and Leading Arm complexes as they are unlikely to accrete within the Gyr.  We calculate their mass as outlined in \S\ref{sec:phys}, which includes the clouds' HI, He, and ionized gas detected in H$\alpha$, and use the angular position of each complex's flux weighted center.  The 23\% of the HVCs (by flux) that do not have a direct distance constraint are not included here, but this gas would not substantially increase the calculated accretion rates.
We ascribe a radial velocity, $v_{r, obs}$, to the complex equal to its flux-weighted radial velocity in GSR.  We then construct a very simple model for the 3-space velocity of the system of all HVC complexes. All complexes are given an identical azimuthal velocity, $v_\theta$, and accretion velocity toward the Galactic center, $v_{G}$. Then, for a given $v_\theta$ and $v_G$, each complex has some minimum additional velocity to make up the difference between the observed radial velocity and modeled radial velocity. The best fit parameters for the model are those that minimize these differences. We note that our results are qualitatively unchanged if the accretion vector is assumed to be toward the disk instead of the Galactic Center.

\begin{figure}[h]
\includegraphics[trim=80 250 0 100, clip=true, scale=0.85]{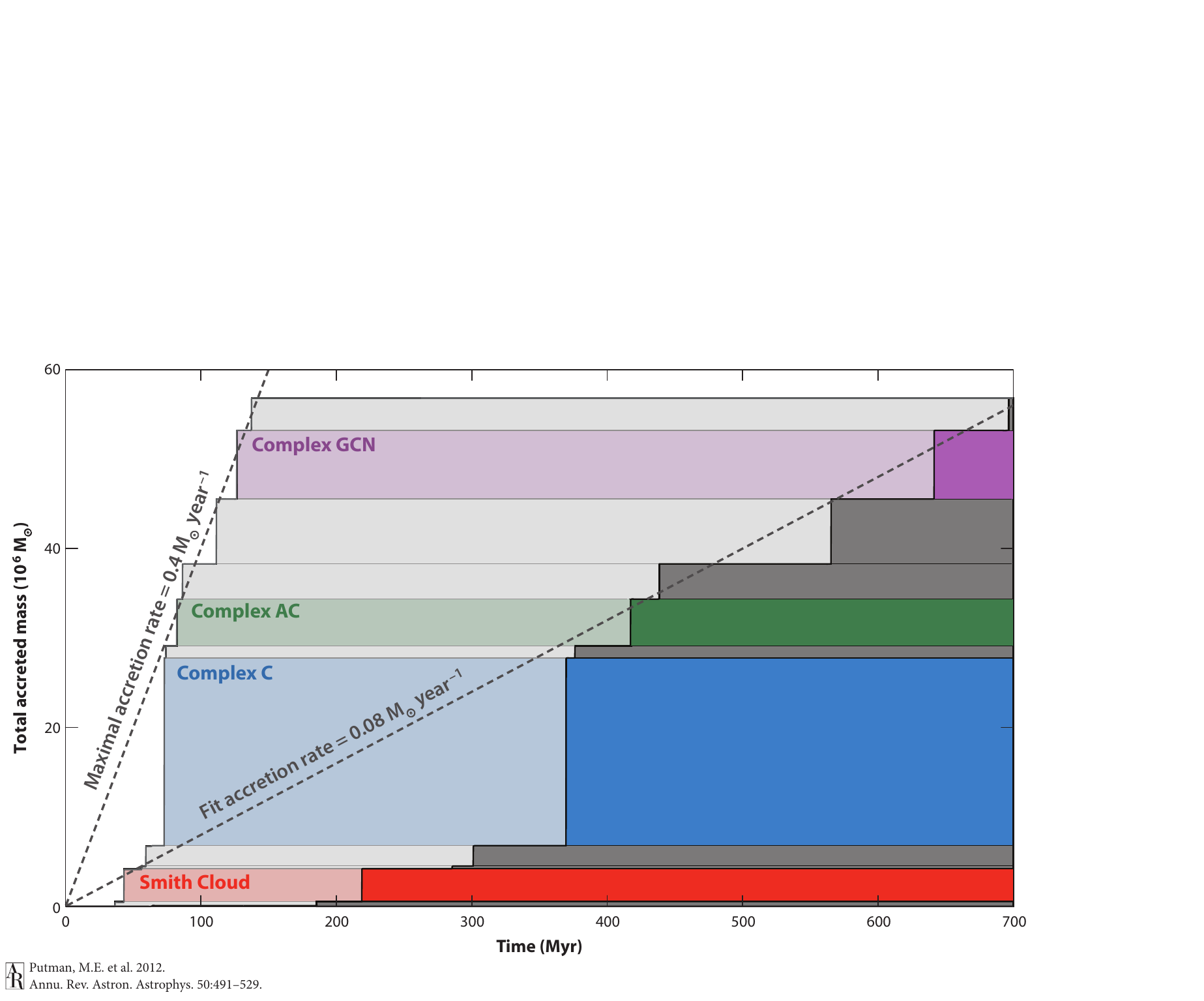}
\caption{\scriptsize The accretion of HVC complex mass with time for the fit (dark shades) and maximal (light shades) accretion velocities $v_G$. Dashed lines indicate accretion rates of 0.08 $M_\odot$/yr and 0.4 $M_\odot$/yr, respectively.  Several specific HVC complexes are noted for reference.   The gap at recent times may be due to selection effects for HVCs close to the disk.}
\label{fig:accrete}
\end{figure}

The complexes are best fit by $[v_\theta, v_G]$ = [77, -40] \kms. We note that the azimuthal velocity, 77\kms~in the direction of Galactic rotation, is roughly consistent with the $v_r$ gradient of -20\kms~kpc$^{-1}$ above the disk noted in \S \ref{sec:diskhalo}, and that the accretion velocity of -40\kms~is roughly consistent with the average radial velocity of all HVCs ($\sim$ -50\kms). We show the accretion of complexes as a function of time given this best fit velocity in Figure \ref{fig:accrete}, as well as a `maximal' accretion rate using the highest $v_G$ allowable with twice the residual in the fit, -200~\kms. The fit and maximal accretion rates are 0.08 and 0.4 $M_\odot$/yr, respectively. The maximal value approaches that required by chemical evolution models for the Milky Way \citep[at least 0.45 $M_\odot$/yr;][]{chiappini09}, but the fit is far below that required.  These values are consistent with accretion rates of neutral gas from simulations \citep{peek08, fernandez12}, and the Complex C accretion model in \cite{thom08}.

Based on Figures~\ref{fig:assoc} and \ref{fig:assoc2} and the discussion at the end of \S\ref{sec:warm}, the ionized component of the HVCs extends beyond the immediate vicinity of the HI component.  The calculation above includes only a factor of two on the HI mass for the warm ionized component directly around HVCs detected in H$\alpha$ emission.  Observations of ionized HVCs in absorption find an additional mass in more diffuse, warm material on the order of $10^8$ \Msun~\citep{lehner11}.   The ionized HVCs are at similar distances to the HI HVCs and also generally have a similar kinematic and spatial distribution, so to roughly account for the diffuse ionized gas the above best fit accretion rate can be scaled up by approximately a factor of two.   Even with this additional ionized component, the accretion rate is still well below that needed by our Galaxy to maintain its current SFR over the next few Gyrs.  Our Galaxy may go through a low point in its gas content, and/or decrease its SFR until the arrival of the gas from the Magellanic System. 

For other galaxies, the amount of HI halo gas is generally between $10^{7.5-9}$ \Msun, but this includes a combination of halo and disk-halo gas (within a few kpc).  Some galaxies show very limited amounts of halo gas and substantial amounts at the disk-halo interface \citep[e.g.,][]{zschaechner11, heald11}.   If accretion rates are estimated from the gas most akin to HVCs, \cite{sancisi08} calculates values of 0.1-0.2 $M_\odot$/yr for several galaxies from the HI component alone.   Large covering fractions of ionized gas are detected through absorption line observations, and so more mass is certainly present in warm and warm-hot halo gas similar to the Milky Way.   Including similar percentages of ionized gas as found for the Milky Way still leaves the accretion rates low compared to each galaxies SFR.   The sources discussed below will provide additional fuel, but low redshift spiral galaxies may also gradually decrease their SFR's \citep[e.g.,][]{madau98,hopkins04}.

\subsection{Disk-Halo Interface}
\label{sec:diskhalofeed}

Halo gas sources need to cool and integrate into the disk in order to become star formation fuel.  Cooling and accretion at the disk-halo interface would be difficult to observe for galaxies beyond $z\sim0$ (i.e., no significant spatial and kinematic offset from the galaxy), and can be called 'quiet accretion'. 
This quiet accretion could explain observations at higher redshifts that show substantial SFRs and outflows with little, if any, evidence for accretion \citep[e.g.,][]{erb08,shapley11,steidel10}.  Cooling at the disk-halo interface also potentially explains the relatively constant amount of HI in the universe since z=3, despite the ongoing star formation \citep[e.g.,][]{prochaska09, putman09w, hopkins08}.   Finally, quiet accretion may be linked to the fact that HI-rich galaxies show quiescent disks \citep{wang11}.

There is observational evidence for infalling structures at the disk-halo interface at low redshift.   
There are large HI features directly connected to galaxy disks in position-velocity space and the IVCs of the Milky Way have a clear bias towards negative velocities.  In addition, the WIM layer of our Galaxy detected in H$\alpha$ shows a net inflow as one examines the regions toward both poles \citep{haffner03, putman09c}.    
A combination of infalling fuel and feedback can successfully model the lagging rotation with z-height found for numerous galaxies \citep{fraternali08, marinacci11}.
The gradient of increasing temperature with increasing z-height (\S\ref{sec:warm} \& \S\ref{sec:otherwarm}) may also  be a combination of the gradual cooling of halo gas, and the hot gas from the disk rising to larger z-heights than the cool gas.
Given stellar metallicity constraints, the ratio of feedback material to fresh gas throughout time requires a full assessment.

The interaction of inflowing clouds with existing halo gas is likely to be important for cooling to occur at the disk-halo interface \citep{marinacci10, heitsch09}.   As clouds move through the halo medium, they are gradually disrupted and slowed.   In simulations of \cite{heitsch09}, the leftover warm over-densities of the HI clouds become buoyant as they approach the disk and are able to re-cool as they are compressed by the surrounding disk-halo medium.  The small clouds that form in this process should be moving similarly to the gas surrounding it at this stage and may be the small cold clouds detected throughout the disk-halo interface of the Galaxy (\S\ref{sec:diskhalo}).  When the clouds begin to fall towards the disk again they will be moving through the surrounding diffuse medium and are likely to resemble the small warm intermediate velocity clouds that show net infall \citep{saul12}, or the filaments/worms that emanate from the HI disk gas  \citep[e.g.,][]{kang07}.

The rate of accretion of different gaseous components at the disk-halo interface can be measured through a comparison of observations to high-resolution simulations.   Preliminary examination of the MW simulation shows a significant amount of accretion of multiphase gas at this interface (\cite{joung12b}; see also \cite{murante12}).  This may indicate that quiet accretion at the disk-halo interface is an important fueling source for spiral galaxy disks.

\subsection{Disk Sources}

Obviously the gas in a galaxy's disk fuels its star formation, however there are two particular fuel reservoirs in the disk for which there is much left to learn.   The first is the gas fed back into the ISM from evolved stars, and the second is outer disk gas that is transported to the inner star forming regions.   
 AGB stars lose on the order of $30-50$\% of their mass back to the ISM according to stellar evolution models \citep{wachter02}. These evolved stars have been observed feeding star formation fuel back into the ISM with CO \citep[e.g.,][]{neri98,castro10} and HI observations \citep[e.g.,][]{gerard06, matthews07, putman11b}.
\cite{leitner10} propose this type of stellar feedback may have provided a significant fraction of a spiral galaxy's fuel over the past few Gyrs.   Galactic stellar metallicities and the deuterium abundance require the feedback material is supplemented with a source of lower metallicity gas.  Understanding the methods with which the feedback material is fed back into the ISM, and the quantity, will help to determine the amount we need from external sources to maintain star formation.  

Another important question related to disk sources of fuel, is how material from gas-rich outer disks is transported to the inner star forming regions.   A large reservoir of fuel is in the outer parts of the disk where few stars are currently forming \citep{werk10b,thilker07}, and the flat metallicity gradients in gas disks indicate transport mechanisms are active \citep{werk10, werk11, bresolin09, spitoni11}. The radial flow of gas in the disk would transport this fuel \citep{sellwood02, elson11}, as well as the potential return of material from a galaxy's warp to more central regions \citep{putman09, jozsa11, kawata03}.   
Perturbations by companions will result in mixing \citep[e.g.,][]{torrey12} and can warp the gas disks \citep{binney92}.   This will also create halo and disk-halo gas features and may be an important part of cycling accreted material inward.

\section{Future Directions}
\label{sec:future}

We have begun to create maps of the multiphase gas in galaxy halos and determine its multiple origins and evolution.   The relationship between halo gas and galaxy feedback and accretion mechanisms can be further clarified through additional studies in several areas.  
For instance, satellite gas is stripped into galaxy halos, but the percentage of halo gas that can be attributed to minor (or potentially past major) mergers remains to be determined.   The IGM is rich with baryons, particularly near star forming galaxies, but our understanding of how the IGM accretes onto galaxies is heavily based on simulation results.   Feedback mechanisms impact galaxy halos, though how effective they are at transporting mass out of the disk and the primary mechanism responsible remains uncertain.  
Finally, although spiral galaxies have abundant fueling resources, the continuity of these resources and the physical mechanisms that lead to the conversion of the incoming fuel into molecular gas have yet to be fully determined.

On the observational front, the next generation of telescopes and instruments will greatly aid our understanding of the flow of baryons in the Universe.   Our knowledge of the link between the cold and warm gas detected in emission and absorption will become stronger with future HI maps with the Square Kilometer Array and the SKA pathfinders.  Three-dimensional maps of the warm gas could be made with large area, high kinematic resolution integral field units (IFUs), and this would show the flow of warm gas relative to the disk.   
The hot halo gas remains an elusive component beyond the disk-halo region.   Highly sensitive x-ray telescopes are needed for a direct detection, but in the meantime we can infer the presence of hot gas with broad warm-hot absorbers in the ultraviolet \citep[e.g.,][]{savage11}.   The motion of the diffuse, hot halo component can potentially be calculated from the properties of the cold and warm clouds embedded within it.
The characterization of the multi-phase halo gas in relation to a wide range of spiral galaxy properties (e.g., total and baryonic mass, SFR, environment) will provide significant insight into its origin.

Stars in galaxy halos will be mapped with future large optical surveys, such as PanSTARRS and LSST.  This will provide more information on the satellite accretion history and stars with a standard intrinsic color can be used for dust detection in the Galactic halo through reddening effects.   Observing the halo stars for absorption lines from halo gas in the optical and ultraviolet will provide tighter distance constraints, and the lines can be used to examine the metallicity and ionization conditions of the halo gas.  With additional estimates of the metallicity, halo gas origin scenarios can be further constrained (in particular the level of feedback), and the dust can be characterized.

On the theoretical front, two of the most important steps are to continue to improve the resolution of the simulations and to study the effect of the many relevant physical processes, both separately and in combination, in a cosmological setting.   Current cosmological models are unable to reach the length scale of many of the observed small HI clouds in the halo and at the disk-halo interface, nor can they accurately resolve the interaction of the multiple phases of gas with one another.  
These problems can be partially alleviated in the near future with models that are limited to a small region in space and time; however, then the relationship of the halo gas to galactic and cosmological processes on large scales is lost.   Until supercomputers become fast enough to cover the large dynamic range required, clever combinations of cosmological and local models can be used to interpret the broad range of observed halo properties and distinguish between various feedback and fueling mechanisms.  Reproducing stellar metallicity distributions with different accretion scenarios will also be an important feature of future simulations.   The current chemical evolution models of our Galaxy are largely semi-analytic \citep[although see][]{few12, kobayashi11}.   Future state-of-the-art simulations can aim to track the metal flows in a galaxy, and through comparisons of observed and simulated stellar metallicities we can learn how feedback and accretion operate throughout time.

\section{Acknowledgements}
We thank our colleagues at Columbia for many useful discussions, in particular comments from Greg Bryan and Jacqueline van Gorkom.
M. Putman thanks several hosts while working on the review, the International Centre for Radio Astronomy Research, the University of Sydney, and Arecibo Observatory.  We thank George Heald, Tom Oosterloo, and Tobias Westmeier for providing HI data, Bart Wakker for providing HVC distance information and funding from NSF grants AST-1008134, AST-0904059 and the Luce Foundation.  JEGP was supported by HST-HF-51295.01A, provided by NASA through a Hubble Fellowship grant from STScI, which is operated by AURA under NASA contract NAS5-26555.


\bibliography{ref}
\bibliographystyle{apj}

\end{document}